\documentclass{aa}
\input{psfig.sty}
\begin{document}
%
%
\title{Radial velocities of early-type stars in the Perseus OB2
association\thanks{Based on observations made at the Observatoire de
Haute Provence (CNRS), France.}}

\author{K.C.\ Steenbrugge\inst{\ 1,2}, J.H.J.\ de Bruijne\inst{\ 1,3},
R.\ Hoogerwerf\inst{\ 1,4} \and P.T.\ de Zeeuw\inst{\ 1}}
\authorrunning{K.C.\ Steenbrugge et al.}

\offprints{K.C.Steenbrugge@sron.nl}

\institute{%
Sterrewacht Leiden, Postbus 9513, 2300 RA Leiden, the Netherlands \and 
Now at SRON National Institute for Space Research, Sorbonnelaan 2, 
3584 CA Utrecht, the Netherlands \and 
Now at Astrophysics Missions Division, European Space Agency, ESTEC, Postbus
299, 2200 AG Noordwijk, the Netherlands \and 
Now at Harvard--Smithsonian Center for Astrophysics, 60 Garden Street,
MS 31, Cambridge, MA 02138, USA}


\hyphenation{Bruij-ne Hoog-er-werf}

\abstract{We present radial velocities for 29 B- and A-type stars in
the field of the nearby association Perseus OB2. The velocities are
derived from spectra obtained with AURELIE, via cross correlation with
radial velocity standards matched as closely as possible in spectral
type. The resulting accuracy is $\sim$$2$--$3$~km~s$^{-1}$. We use
these measurements, together with published values for a few other
early-type stars, to study membership of the association. The mean
radial velocity (and measured velocity dispersion) of Per~OB2 is $23.5
\pm 3.9$~km~s$^{-1}$, and lies $\sim$$15$~km~s$^{-1}$ away from the
mean velocity of the local disk field stars. We identify a number of
interlopers in the list of possible late-B- and A-type members which
was based on Hipparcos parallaxes and proper motions, and discuss the
colour-magnitude diagram of the association.
}

\maketitle

\section{Introduction} 
\label{sec:intro} 
 
Perseus OB2 is one of the nearest OB associations. It was discovered
visually as a loose group of 15 bright, blue stars by Blaauw (1944),
who subsequently confirmed his finding by means of radial velocity
data from Moore's (1932) catalogue. Blaauw (1952; B52; see also Blaauw
1964) established membership for 17 bright O and B stars based on
proper motion data. These include a number of well-studied
spectroscopic binaries (e.g., AG~Per), as well as the high-mass X-ray
binary X~Per.
 
De Zeeuw et al.\ (1999; Z99) published an updated member list for Per
OB2, based on Hipparcos position, proper motion, and parallax data,
containing 17 B- and 16 A-type stars plus a small number of late-type
stars. Based on extensive modeling of the kinematics of the Galactic
disk, taking selection criteria in the Hipparcos Catalogue into
account, Z99 concluded that a significant fraction of the A- and
later-type stars identified as astrometric members are likely to be
interlopers. This conclusion was confirmed by Belikov et al.\ (2002a,
b), who used Tycho--2 proper motions as well as photometric
information. These authors also identified nearly 1000 additional
probable members of Per~OB2 to 12th magnitude, and suggested that this
association may consist of two subgroups.

\begin{figure*}[t!] 
\centerline{\psfig{file=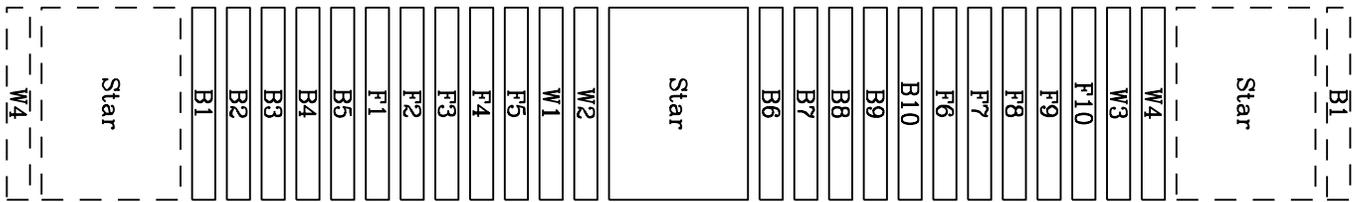,width=\textwidth,silent=,clip=}} 
\caption[]{Observing strategy. Each object exposure (either target or
standard star) was preceded by a calibration sequence, consisting of
five bias exposures (B1 through B5), five flat field exposures (F1
through F5), and two wavelength calibration arc spectra (W1 and W2),
and followed by an identical calibration sequence (B6 through B10, F6
through F10, W3 and W4). In order to minimize overhead, we used the
trailing calibration sequence of each star also as the leading
sequence for the next star. Time progresses from left to right.}
\label{fig:1} 
\end{figure*}

The Hipparcos satellite (ESA 1997) measured stellar positions,
parallaxes, and proper motions, providing five-dimensional data in
phase space. The sixth component, the radial velocity, is important,
e.g., for improving membership and expansion studies (e.g., Brown,
Dekker \& de Zeeuw 1997). This is particularly true for Per~OB2, as
the space motion of the association relative to the local disk
population is mostly along the line-of-sight and proper motions of
member stars are consequently small. Unfortunately, a homogeneous set
of radial velocities is not available for the (early-type) Hipparcos
members of Per~OB2. In this paper, we present new spectroscopic
observations of 29 B- and A-type stars in Per~OB2.
 
While determining (relative) spectroscopic radial velocities for
late-type stars using cross correlation techniques is feasible with
precisions of tens of m~s$^{-1}$, substantial effort is required to
achieve even km~s$^{-1}$-level precisions for early-type stars (e.g.,
Verschueren \& David 1999; Verschueren, David \& Griffin 1999;
Griffin, David \& Verschueren 2000). Early-type optical spectra show
few absorption lines and these lines are intrinsically broad (up to a
few hundred km~s$^{-1}$). They are often broadened even further by
stellar rotation and sometimes also show variability due to pulsations
and/or stellar winds. The resulting correlation peaks are broad, and
can have significant substructure caused by the mixing of spectral
lines (cf.\ figure~1 in Verschueren \& David 1999). These effects
complicate accurate centering of correlation peaks and thus the
precise determination of (even relative) radial velocities. The
derived radial velocities of OBA-type stars, moreover, are known to
depend on the spectral region used in the cross correlation (e.g.,
Verschueren et al.\ 1999). Furthermore, these stars emit their
radiation primarily in the blue part of the spectrum, where CCD
devices have reduced quantum efficiencies and slit-centering and
atmospheric refraction errors are potentially significant (e.g.,
Verschueren et al.\ 1997). Finally, unrecognized multiplicity can be a
significant source of error.

We show here that, despite these significant complications,
high-resolution spectroscopy combined with a careful observing
strategy which includes measurement of many standard stars, does allow
measurement of radial velocities for B and A-type stars with an
accuracy of a few km~s$^{-1}$. This makes it possible to improve
substantially the membership list for the early-type members of
Per~OB2. We describe our observations in Sect.~2. In Sect.~3, we focus
on the data reduction. The data analysis is presented in Sect.~4. The
interpretation of the results and the conclusions are given in
Sections~5 and 6, respectively.
 
\begin{figure*}[t!] 
\centerline{\psfig{file=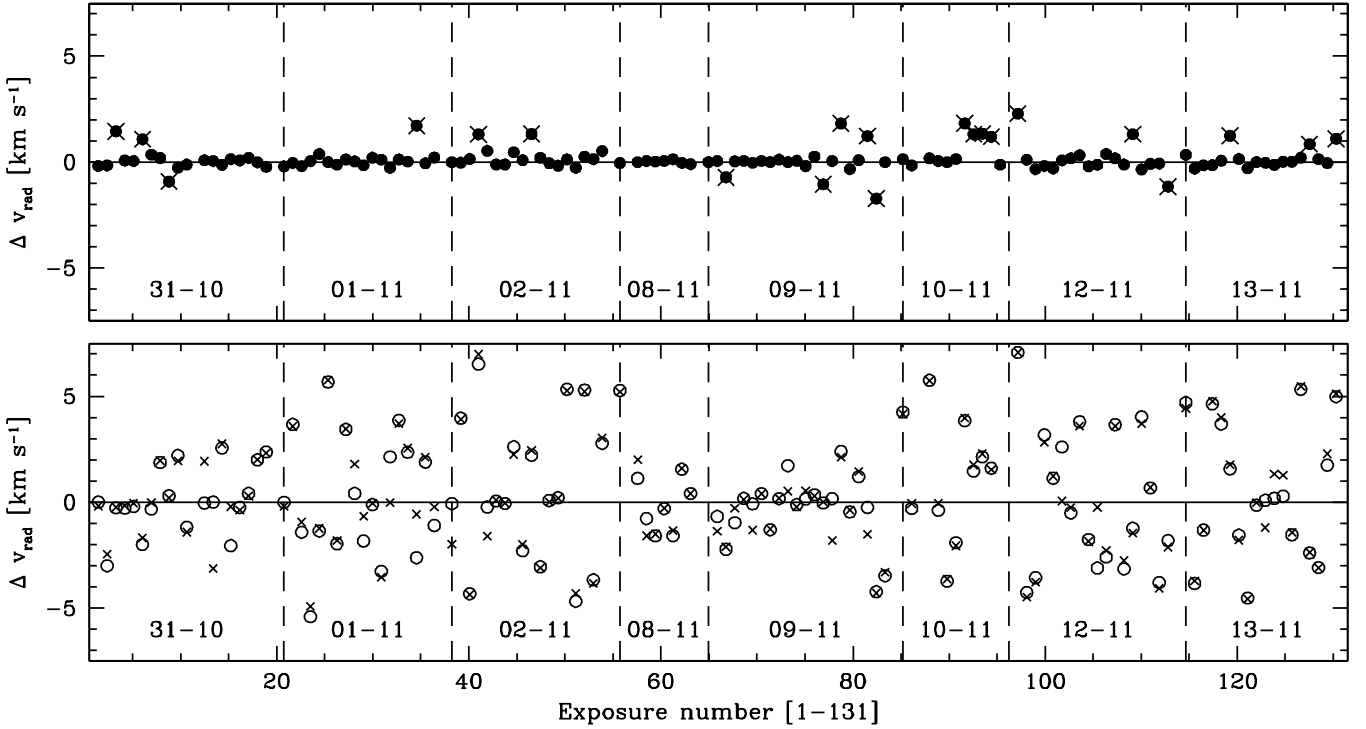,width=\textwidth,silent=}} 
\caption[]{The stability of the detector, AURELIE. Each of the 131 
star exposures has four associated wavelength calibration lamp spectra
(W1 through W4): W1 and W2 were obtained before and W3 and W4 after
each stellar exposure (Figure~\ref{fig:1}). Lamp spectra W3 and W4 were
generally taken $\sim$30--45~min later than W1 and W2. {\it Top
panel:\/} the filled circles show the shift (in km~s$^{-1}$) between
the velocity scales inferred from W1 and from W2. This shift is, for
our purposes, acceptably small ($\sigma = 0.6$~km~s$^{-1}$), with the
exception of 21 cases (crossed symbols). These exposures are treated with
extra care in this study. The dashed vertical lines separate different
observing nights, the start dates of which are printed in the panel
(year: 1997). {\it Bottom panel:\/} open circles and crosses denote
the shifts between W1 and W3 and W1 and W4, respectively. Random shifts
up to $\sim$10~km~s$^{-1}$ occur ($\sigma = 2.9$~km~s$^{-1}$), most
likely due to instability of the detector (Gillet et al.\ 1994).}
\label{fig:2} 
\end{figure*}

\section{Observations} 
\label{sec:obs} 
 
The spectra presented here were obtained between Oct.\ 31 and Nov.\
13, 1997, with the 1.52~m Coud\'{e} telescope at Observatoire de
Haute--Provence (OHP). We used the AURELIE detector, a linear
2048-pixel photodiode array, which is relatively blue-sensitive
(Gillet et al.\ 1994). We used grating R3 with a linear dispersion of
$\sim$16.5~\AA~mm$^{-1}$ and a resolving power $R \sim 7000$ in order
to achieve a precision of a few km~s$^{-1}$ in the final radial
velocities; this precision is of the same order of magnitude as the
typical error in tangential space motions due to Hipparcos proper
motion errors at the distance of the Per~OB2 association ($318 \pm
27$~pc; Z99), and comparable to the expected internal velocity
dispersion of the group. We used a Th--Ar lamp for wavelength
calibration, as well as an image slicer with a circular
$3^{\prime\prime}$ entrance aperture, which minimizes potential
wavelength errors due to poor centering (Gillet et al.\ 1994). As our
targets are hot B- and A-stars, which primarily show strong hydrogen
and helium lines, we used the spectral range from 3800 to
4200~\AA. This choice resulted in the presence of about 10
higher-order Balmer and helium lines in a typical spectrum. The
$3800$--$4200$~\AA\ spectral region also contains the
Ca{\small{\sc{II}}} H- and K-lines (K: $\lambda = 3933$~\AA; H:
$\lambda = 3968$~\AA; cf.\ Figure~\ref{fig:3}). In B-type stars, the
K-line is mostly interstellar, but at spectral types around A0 and
later, this line, together with other metallic lines, becomes a
notable feature intrinsic to the stellar spectra. For early-type
stars, the Ca{\small{\sc{II}}} H- and K-lines thus trace gas along the
line of sight instead of stellar kinematics (see, e.g., Sonnentrucker
et al.\ 1999, who actually studied the gas and dust distribution
towards Per~OB2 using high-resolution spectra of bright stars in the
association). Although the presence of (Doppler-shifted) H- and
K-lines can potentially affect cross correlation results and the
associated radial velocities (Sect.~\ref{sec:data_analysis}), the
Ca{\small{\sc{II}}} metal lines carry minimal weight in practice and
their effects are negligibly small.
 
Z99 established membership of Per~OB2 by applying two independent
selection methods (the convergent point method of de Bruijne [1999]
and the Spaghetti method of Hoogerwerf \& Aguilar [1999]) to all
Hipparcos entries in a specific field on the sky (see table~A1 in
Z99). This approach classified each star as either `certain member'
(acceptance by both methods), `possible member' (acceptance by one of
the two methods exclusively), or `non-member' (rejection by both
methods). As targets for our observations, we selected the 33 B- and
A-type certain members of Per~OB2 identified by Z99. We added a number
of their possible members, as well as some early-type members from B52
which were not confirmed by Z99.
 
We observed the resulting list in order of decreasing brightness, 
ultimately obtaining high-quality spectra for 29 distinct targets 
(Table~\ref{tab:1}). Exposure times ranged from 10 to 30~min. Many 
targets were observed multiple times. In total, more than 7 nights of 
our 2-week observing campaign were weathered out completely. 
 
In addition, we repeatedly obtained high-quality spectra for 20
(candidate) radial-velocity standard stars. As an IAU-approved list of
early-type radial velocity standards does not exist, we selected these
stars from various sources (Table~\ref{tab:2};
Sect.~\ref{subsec:standard_stars}). In this selection, we tried to
cover spectral type and luminosity class ranges as large as
possible. We also preferentially selected stars with small rotation
velocities. Although time consuming, the necessity to build up such a
private standard star library, observed with the same instrumental
setup as used for the target stars, has the advantage of working with
a homogeneous data set, and optimizes the accuracy of the final
results (we use `precision' for random and `accuracy' for systematic
errors).
 
We obtained a total of 131 object exposures (target and standard stars 
together). We decided to take a calibration sequence, consisting of 
five bias exposures, five flat fields, and two wavelength calibrations, 
before and after {\it every\/} star was observed (Figure~\ref{fig:1}). 
In order to reduce overhead, we re-used the trailing calibration 
sequence of each star as the leading sequence for the next star to be 
observed. Although this procedure still resulted in a significant 
overhead, it helped to minimize the effect of instrumental errors in 
our final results.

\renewcommand{\tabcolsep}{6.0pt} 
\begin{table*}[th!] 
\caption[]{Observed target stars: B- and A-type members of Per~OB2 
identified by Z99, supplemented with a number of early-type stars 
identified as members in pre-Hipparcos investigations (e.g., B52). The 
stars were observed in order of decreasing brightness, resulting in 
spectra for 29 distinct targets (62 exposures). Columns: 
(1) HD number (Hipparcos Catalogue field H71); 
(2) Hipparcos identifier (H1); 
(3) name; 
(4) $V$ magnitude (H5); 
(5) spectral type and luminosity class;
(6) number of spectra obtained ($N$; subtractions refer to suspect
    exposures related to detector instability); 
(7) multiplicity (SB: spectroscopic binary; C: Hipparcos component
    binary --- $\rho$ and $\Delta Hp$ in column~11; G: Hipparcos
    acceleration binary; S: Hipparcos suspected non-single --- field H61); 
(8) Hipparcos membership (C: Z99 certain member; P: Z99 possible member;
    B: Z99 non-member, but B52 member); 
(9) Hipparcos parallax $\pi$ (H11; mas);
(10) Hipparcos parallax error $\sigma_\pi$ (H16; mas);
(11) Hipparcos components. 
}
\begin{center} 
\begin{tabular}{rcccllccccl}
\hline\hline 
\multicolumn{1}{c}{HD} & 
\multicolumn{1}{c}{HIP} & 
\multicolumn{1}{c}{Name} & 
\multicolumn{1}{c}{$V$} & 
\multicolumn{1}{c}{SpT+LCl} & 
\multicolumn{1}{c}{$N$} & 
\multicolumn{1}{c}{Mult.} & 
\multicolumn{1}{c}{Z99} & 
\multicolumn{1}{c}{$\pi$} & 
\multicolumn{1}{c}{$\sigma_\pi$} & 
\multicolumn{1}{l}{Hipparcos components}\\
\multicolumn{1}{c}{(1)} & 
\multicolumn{1}{c}{(2)} & 
\multicolumn{1}{c}{(3)} & 
\multicolumn{1}{c}{(4)} & 
\multicolumn{1}{c}{(5)} & 
\multicolumn{1}{c}{(6)} & 
\multicolumn{1}{c}{(7)} & 
\multicolumn{1}{c}{(8)} & 
\multicolumn{1}{c}{(9)} & 
\multicolumn{1}{c}{(10)} & 
\multicolumn{1}{l}{(11)} \\
\hline 
 18830 & 14207 &             & 8.34 & A0       & $3  $ & C    & C &  2.50& 1.42& $\rho=0{\farcs}335, \Delta Hp=2.27$~mag\\ 
 19216 & 14450 &             & 7.84 & B9V      & $3  $ &      & P &  4.81& 1.01& \\ 
 19567 & 14713 &             & 7.62 & B9       & $2-1$ &      & C &  4.54& 0.96& \\ 
 20113 & 15151 &             & 7.65 & B8       & $4-1$ &      & P &  3.43& 0.95& \\ 
 20987 & 15895 &             & 7.87 & B2V      & $2  $ & G    & C &  1.80& 1.08& \\ 
 21483 & 16203 &             & 7.06 & B3III    & $2-1$ &      & B &  1.60& 1.05& \\ 
 21856 & 16518 &             & 5.91 & B1V      & $1  $ &      & B &  1.99& 0.82& \\ 
 22114 & 16724 &             & 7.60 & B8Vp     & $2-1$ &      & P &  3.62& 0.99& \\ 
 22951 & 17313 & 40 Per      & 4.97 & B0.5V    & $1  $ & S    & C &  3.53& 0.88& \\ 
 23060 & 17387 &             & 7.51 & B2Vp     & $2  $ &      & P &  2.09& 0.93& \\ 
 23180 & 17448 & o~Per       & 3.84 & B1III    & $3-1$ & C+SB & P &  2.21& 0.84& $\rho=1{\farcs}019, \Delta Hp=2.91$~mag\\ 
 23268 & 17498 &             & 8.22 & A0       & $3  $ &      & C &  3.34& 0.99& \\ 
 23478 & 17631 &             & 6.68 & B3IV...  & $1  $ &      & B &  4.19& 1.03& \\ 
 23597 & 17698 &             & 8.19 & B8       & $3-1$ &      & C &  2.56& 1.01& \\ 
 23625 & 17735 &             & 6.57 & B2.5V    & $1  $ & C+SB & C &  2.63& 1.00& $\rho=3{\farcs}349, \Delta Hp=2.58$~mag\\ 
 23802 & 17845 &             & 7.45 & B5Vn     & $3  $ & C    & C &  3.09& 1.21& $\rho=2{\farcs}240, \Delta Hp=2.32$~mag\\ 
 24012 & 17998 &             & 7.84 & B5       & $2  $ &      & C &  1.82& 1.12& \\ 
 24131 & 18081 &             & 5.78 & B1V      & $2-2$ &      & C &  3.15& 0.84& \\ 
 24190 & 18111 &             & 7.43 & B2V      & $2-1$ & SB   & C &  2.04& 1.00& \\ 
 24398 & 18246 & $\zeta$~Per & 2.84 & B1Ib     & $1  $ &      & C &  3.32& 0.75& \\ 
 24583 & 18390 &             & 9.00 & B7V      & $5  $ &      & P &  3.31& 1.35& \\ 
 24640 & 18434 &             & 5.49 & B1.5V    & $1  $ & S    & P &  3.36& 0.76& \\ 
 24970 & 18621 &             & 7.44 & A0       & $3-1$ &      & P &  4.95& 0.99& \\ 
 25539 & 19039 &             & 6.87 & B3V      & $1  $ &      & C &  4.19& 0.97& \\ 
 25799 & 19178 &             & 7.05 & B3V...   & $2  $ & SB   & C &  2.78& 0.95& \\ 
 25833 & 19201 & AG~Per      & 6.70 & B5V:p    & $1-1$ & C+SB & C &  3.89& 1.31& $\rho=0{\farcs}803, \Delta Hp=1.81$~mag\\ 
 26499 & 19659 &             & 9.06 & B9       & $2  $ &      & C &  4.14& 1.30& \\ 
278942 & 17113 &             & 9.03 & B3III    & $2  $ & C    & C &  4.83& 1.21& $\rho=0{\farcs}149, \Delta Hp=0.78$~mag\\ 
281159 & 17465 &             & 8.51 & B5V      & $2  $ & C    & P &  4.52& 3.30& $\rho=0{\farcs}615, \Delta Hp=0.25$~mag\\ 
\hline\hline 
\end{tabular} 
\label{tab:1} 
\end{center} 
\end{table*}
 
\section{Data reduction} 
\label{sec:data_reduction} 
 
The data reduction was performed using IRAF software (Tody 1993). For
each target exposure, we first subtracted the average bias exposure
calculated from bias exposures B1 through B10 (Figure~\ref{fig:1}). In
the next step, we similarly divided by an average flat field
exposure. However, we rejected the first of the five flat field exposures
in each calibration sequence (F1 and F6; Figure~\ref{fig:1}) in order
to allow for warming-up effects of the flat field lamp. We then
manually removed cosmic rays from our data, after which we
wavelength-calibrated the Th--Ar lamp spectra.
 
In order to verify the stability of the detector, we cross correlated
the four lamp spectra W1 through W4 taken before and after each star
was observed. We found generally good agreement between the wavelength
scales inferred from the two lamp spectra taken either before (W1 and
W2) or after (W3 and W4) the target exposure ($\sigma = 0.6$ and
$0.7$~km~s$^{-1}$, respectively; Figures~\ref{fig:1} and
\ref{fig:2}). However, cross correlating W1 with either W3 or W4,
which were taken $\sim$30--45~min later, showed random shifts of up to
$\pm$10~km~s$^{-1}$ ($\sigma = 2.9$~km~s$^{-1}$). These shifts
effectively degrade the resolution of the detector, and are most
likely due to thermal instabilities of the detector, which are known
to be present at the level of $\sim$~0.1~pixel~(hr$^{-1}$) or
$\sim$15~km~s$^{-1}$~(hr$^{-1}$) (Gillet et al.\ 1994). Shifts between
W1/W2 and W3/W4 cannot be due to telescope re-pointings as AURELIE is
located on a dedicated optical bench in the telescope control room. We
did not detect any systematic, long-term trend (spanning hours to
days) in the detector zero point, consistent with the small
repeatability errors derived for a number of late-type stars that were
observed throughout our campaign (Sect.~\ref{subsec:reper}).
 
\begin{figure*}
\centerline{\psfig{file=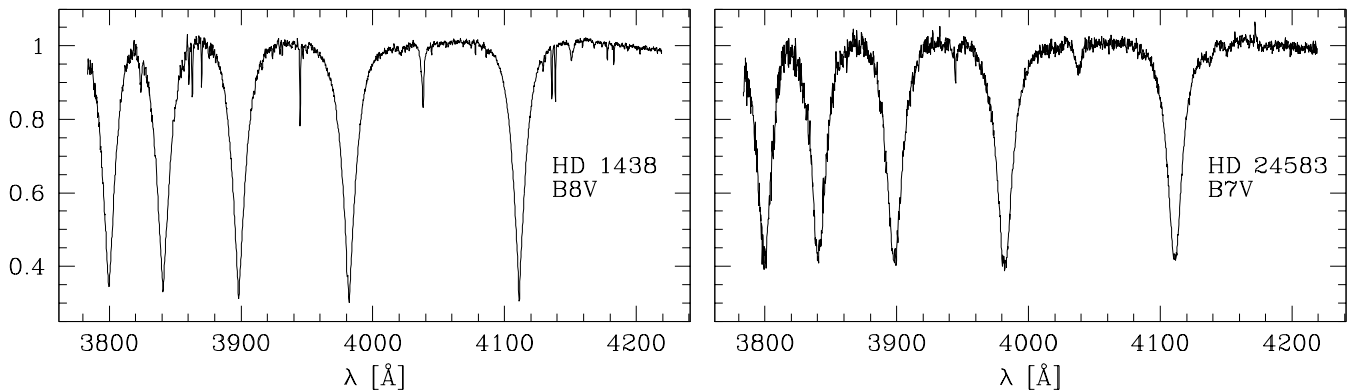,width=\textwidth,silent=,clip=}} 
\caption[] {Example of reduced (continuum-subtracted) spectra:
HD~1438 (B8V; $V = 6.10$~mag; $T = 30$~min; left panel) and HD~24583
(B7V; $V = 9.00$~mag; $T = 30$~min; right panel). Note the
near-absence of metal lines, the broad nature of the higher-order
Balmer hydrogen lines and the helium lines, and the spectral
differences in the helium lines.} \label{fig:3}
\end{figure*} 

Because the lamp spectra in the first calibration sequence (W1 and W2)
were taken immediately before the stellar exposure (Figure~\ref{fig:1}),
we decided not to use the two lamp spectra in the following sequence
(W3 and W4), as these were taken after another five bias exposures and
flat-field exposures. Unfortunately, even W1 and W2, taken within
minutes of each other, show significant shifts ($> \sigma =
0.6$~km~s$^{-1}$) in 21 cases (crossed symbols in
Figure~\ref{fig:2}). We reject the associated exposures in case of a
standard star pointing (10 observations; see column~5 in
Table~\ref{tab:2}) and treat these exposures with special care in case of
a target pointing (11 observations; see column~6 in
Table~\ref{tab:1}).
 
In the final step of the data reduction procedure, we thus wavelength
calibrated our spectra using the W1 Th-Ar lamp spectra. Examples
of two fully reduced (continuum-subtracted) spectra are shown in
Figure~\ref{fig:3}.

\section{Data analysis} 
\label{sec:data_analysis} 
 
Radial velocities are generally determined by means of spectral cross 
correlation between the target and a suitable standard star (but see, 
e.g., Dravins, Lindegren \& Madsen 1999). Standard star spectra can 
either be obtained observationally or generated by means of stellar 
evolutionary codes (e.g., Morse, Mathieu \& Levine 1991). Whereas the 
latter approach has the advantage that the spectral type, luminosity 
class, rotation velocity, etc.\ of the template can be chosen 
representative of the target spectrum, significant uncertainties exist 
in the (atmospheric) models of early-type stars (e.g., Lennon, Dufton 
\& Fitzsimmons 1992). We chose to use our own standard star spectra 
for radial velocity calibration. Possible drawbacks in this approach 
are, e.g., unrecognized multiplicity and errors in catalogued radial 
velocities and/or spectral classifications.

\renewcommand{\tabcolsep}{6.7pt} 
\begin{table*}[th!] 
\caption[]{Stars used as radial velocity standards. Columns: 
(1) HD number (Hipparcos Catalogue field H71); 
(2) Hipparcos identifier (H1); 
(3) $V$ magnitude (H5); 
(4) spectral type and luminosity class (H76); 
(5) number of spectra obtained ($N$; subtractions refer to rejected
    exposures due to detector instability); 
(6) multiplicity (G: Hipparcos acceleration binary; S: Hipparcos
    suspected non-single --- field H61)
(7) assumed radial velocity (km~s$^{-1}$; see
    Sect.~\ref{subsec:standard_stars} for details);
(8) idem from Morse et al.\ (1991; table 6; table 4 for HD~23408); 
(9) radial velocity in km~s$^{-1}$ from Fekel (1985; table 3);
(10) idem from Wolff (1978; ${\rm range} < 3 \cdot {\rm
     standard~deviation}$; tables 1 and 3);
(11) idem from Abt \& Levy (1978; table 1); 
(12) idem from Gies \& Bolton (1986; table 3); 
(13) idem from Latham \& Stefanik (1982; table 1).} 
\begin{center} 
\begin{tabular}{rrcllcrrrrrrr} 
\hline\hline 
\multicolumn{1}{c}{HD} &
\multicolumn{1}{c}{HIP} &
\multicolumn{1}{c}{$V$} &
\multicolumn{1}{c}{SpT+LCl} &
\multicolumn{1}{c}{$N$} &
\multicolumn{1}{c}{Mult.} &
\multicolumn{7}{c}{Radial velocity [km~s$^{-1}$]}\\
\multicolumn{1}{c}{(1)} &
\multicolumn{1}{c}{(2)} &
\multicolumn{1}{c}{(3)} &
\multicolumn{1}{c}{(4)} &
\multicolumn{1}{c}{(5)} &
\multicolumn{1}{c}{(6)} &
\multicolumn{1}{c}{(7)} &
\multicolumn{1}{c}{(8)} &
\multicolumn{1}{c}{(9)} &
\multicolumn{1}{c}{(10)} &
\multicolumn{1}{c}{(11)} &
\multicolumn{1}{c}{(12)} &
\multicolumn{1}{c}{(13)}\\ 
\hline 
1438   & 1501   & 6.10 & B8V     & $3  $ & &    +3.3\phantom{$^{\rm a}$} & .       & .       & .       & .      & .     & +3.3   \\ 
3360   & 2920   & 3.69 & B2IV    & $3-1$ & &  $-$0.3\phantom{$^{\rm a}$} & $-$0.3  & .       & .       & +1.0   & .     & +0.5   \\ 
10982  & 8387   & 5.86 & B9V     & $3-1$ & &    +5.0\phantom{$^{\rm a}$} & .       & .       & +5.0    & .      & .     & .      \\ 
17081  & 12770  & 4.24 & B7IV    & $2-1$ & &   +15.4\phantom{$^{\rm a}$} & +15.4   & +14.3   & .       & .      & .     & .      \\ 
23408  & 17573  & 3.87 & B8III   & $2  $ & &    +6.5\phantom{$^{\rm a}$} & +6.5    & .       & .       & .      & .     & +7.6   \\ 
26912  & 19860  & 4.27 & B3IV    & $2-1$ & &   +14.9\phantom{$^{\rm a}$} & +14.9   & .       & .       & +15.5  & .     & .      \\ 
27638  & 20430  & 5.38 & B9V     & $2  $ &S&   +14.4\phantom{$^{\rm a}$} & .       & .       & .       & .      & .     & +14.4  \\ 
28114  & 20715  & 6.06 & B6IV    & $2-1$ & &   +12.9\phantom{$^{\rm a}$} & .       & .       & +12.9   & .      & .     & +13.4  \\ 
35708  & 25539  & 4.88 & B2.5IV  & $2  $ & &   +17.0\phantom{$^{\rm a}$} & +17.0   & +18.1   & .       & +15.2  & .     & +18.4  \\ 
36267  & 25813  & 4.20 & B5V     & $3-1$ & &   +19.8\phantom{$^{\rm a}$} & .       & .       & .       & +19.8  &  .    & .      \\ 
38899  & 27511  & 4.89 & B9IV    & $2  $ & &   +21.6\phantom{$^{\rm a}$} & +21.6   & +21.6   & +22.2   & .      & .     & +20.9  \\ 
43112  & 29678  & 5.91 & B1V     & $2  $ & &   +37.3\phantom{$^{\rm a}$} & +37.3   & .       & .       & .      & +35.8 & .      \\ 
58142  & 36145  & 4.61 & A1V     & $5-1$ & &   +26.0\phantom{$^{\rm a}$} & +26.0   & +27.2   & .       & .      & .     & +26.9  \\ 
196724 & 101867 & 4.81 & A0V     & $6-1$ &G& $-$18.4$^{\rm a}$           & $-$12.0 & .       & .       & .      & .     & .      \\ 
196821 & 101919 & 6.08 & A0III   & $4-1$ & & $-$31.3\phantom{$^{\rm a}$} & .       & $-$31.3 & $-$31.6 & .      & .     & .      \\ 
201345 & 104316 & 7.78 & O9p     & $1  $ & &   +21.6\phantom{$^{\rm a}$} & .       & .       & .       & .      & +21.6 & .      \\ 
214994 & 112051 & 4.80 & A1IV    & $5  $ &G&    +9.1\phantom{$^{\rm a}$} & +9.1    & +7.5    & .       & .      & .     & +8.5   \\ 
217811 & 113802 & 6.37 & B2V     & $3  $ & & $-$11.2\phantom{$^{\rm a}$} & $-$11.2 & .       & .       & .      & .     & $-$10.2\\ 
219188 & 114690 & 7.06 & B0.5III & $4-1$ & &   +48.0$^{\rm b}$           & .       & .       & .       & .      & +68.0 & .      \\ 
220599 & 115591 & 5.56 & B9III   & $2  $ & &   +12.0\phantom{$^{\rm a}$} & .       & .       & +12.0   & .      & .     & .      \\ 
\hline\hline 
\end{tabular} 
\end{center} 
\label{tab:2} 
\vskip-0.25truecm
$^{\rm a}$ Fekel 1990 (cf.\ Liu, Janes \& Bania 1989);\hfill\break
$^{\rm b}$ SIMBAD.\hfill\break 
\end{table*}

\subsection{Standard stars} 
\label{subsec:standard_stars} 
 
Our 20 early-type standard stars have been studied carefully in the 
past, mostly with the aim of establishing binarity, and they have 
never shown signs of velocity variability, although HD~27638, 
HD~196724, and HD~214994 may have astrometric companions 
(Table~\ref{tab:2}). Selection criteria in the construction of this 
standard star sample included visibility, visual magnitude, spectral 
type, luminosity class, and small rotational velocity. 
 
The published radial velocities for our standard stars generally agree 
to within $\sim$$1$~km~s$^{-1}$, but differences up to 
$\sim$$3$~km~s$^{-1}$ occur. Some of these discrepancies may be due to 
differences in the absolute zero points used in different studies. In 
order to minimize the effect of this uncertainty, we preferentially 
selected the radial velocities for our standard stars from the source 
containing most measurements in absolute numbers (Morse et al.\ 1991; 
results based on CCD data). For the remaining stars, we used, in order 
of decreasing preference, Fekel (1985), Wolff (1978), Abt \& Levy 
(1978), Gies \& Bolton (1986), and Latham \& Stefanik 
(1982). Exceptions to this rule were made for HD~196724 and 219188, 
for which we assume a radial velocity of $-18.4$~km~s$^{-1}$ (Fekel 
1990; Liu, Janes \& Bania 1989) and $+48.0$~km~s$^{-1}$ (SIMBAD), 
respectively. Upon comparing the accordingly selected radial 
velocities with the values contained in the compilation catalogue of 
mean radial velocities of Barbier--Brossat \& Figon (2000), we find a 
mean difference of $0.1$~km~s$^{-1}$ and a standard deviation of 
$2.0$~km~s$^{-1}$. These values (probably) mostly reflect zero-point 
differences (and random errors) and imply the accuracy and precision 
of our final radial velocities are $\ga$0.1~km~s$^{-1}$ and 
$\ga$2.0~km~s$^{-1}$, respectively.

\subsection{Cross correlation} 
 
We cross correlated the spectra by means of the standard IRAF task
{\tt fxcor}. In order to suppress noise, we applied high-frequency
Fourier filtering to all spectra (cf.\ Verschueren \& David 1999). We
did not apply low-pass filtering, and used a Gaussian fitting function
for centering of the correlation peak. This function was found
empirically to be a suitable choice. The detector instability did not
justify a more refined fitting function. In order to suppress the
effect of continuum mismatch, we subtracted the continuum from both
the template and the target spectrum --- and thus normalized both
spectra to a continuum level of $1$ --- before each cross
correlation.
 
\begin{figure}[t!] 
\centerline{\psfig{file=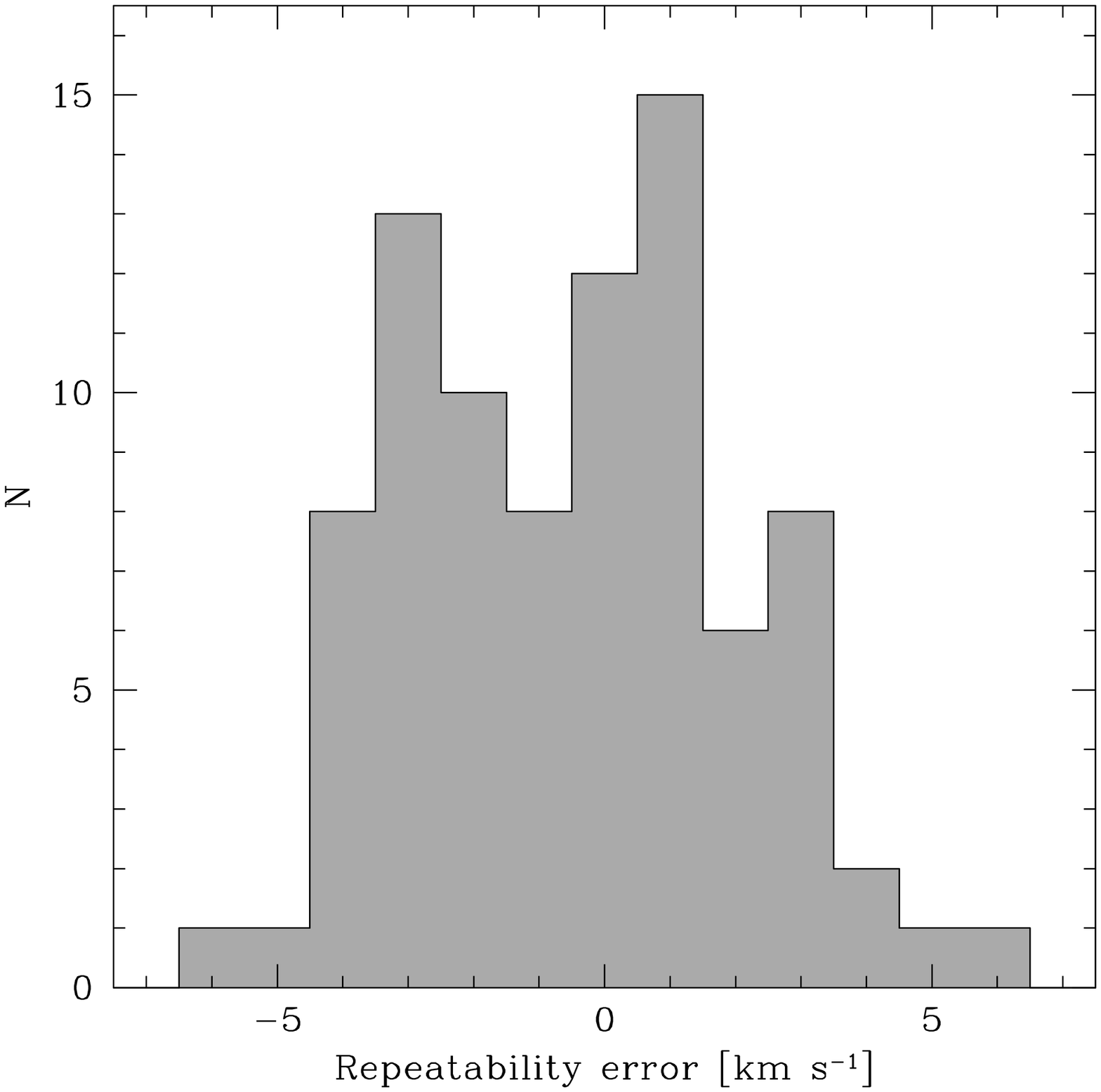,width=8.43truecm,silent=,clip=}} 
\caption[]{Repeatability errors for 86 pairs of spectra (either target
or standard stars; see Sect.~\ref{subsec:reper} for details). In the
remainder of this paper, we assume that repeatability errors
($\sigma_{\rm r}$) are positive. The mean and median repeatability
errors are $2.1$ and $2.2$~km~s$^{-1}$, respectively.}
\label{fig:reper} 
\end{figure}

\subsection{Repeatability errors} 
\label{subsec:reper} 
 
We observed many of the objects, either target or standard stars,
multiple times. Under the assumption that these stars are single,
cross correlation of their spectra (on an object-by-object and
exposure-by-exposure basis) should ideally result in vanishingly small
shifts. Conversely, shifts measured in such an exercise provide a
direct estimate of the random errors that will be present in the final
radial velocities (e.g., Morse et al.\ 1991; Verschueren et al.\
1997). In our case, these random errors include the temporal
instability of the detector (Sect.~\ref{sec:data_reduction}).
 
Figure~\ref{fig:reper} shows the repeatability errors inferred from 
our data, after excluding known spectroscopic binaries. We define the 
repeatability error $\sigma_{\rm r}$ for a pair of exposures of a given 
object as the relative shift (in km~s$^{-1}$) emerging during their 
cross correlation. Sixteen of our standard stars and 18 of our Per~OB2 
targets were observed multiple times, resulting in 86 repeatability 
errors (the two exposures of the faint target HD~278942 lead to a 
repeatability error of $\sim$$16$~km~s$^{-1}$; as these spectra are 
noise-dominated, we reject this object from this discussion). From 
Figure~\ref{fig:reper} we conclude that the repeatability errors are 
at the level of $\sim$$0$--$3$~km~s$^{-1}$, consistent with the 
expected noise due to thermal instabilities of the detector 
(Figure~\ref{fig:2}). The mean repeatability error is 
$\overline{\sigma_{\rm r}} = 2.1$~km~s$^{-1}$ (the median is 
$2.2$~km~s$^{-1}$); this value could be slightly overestimated as a 
result of unrecognized multiplicity. A value of $2.1$~km~s$^{-1}$ is 
comparable to repeatability errors quoted in the literature (e.g., 
Morse et al.\ 1991: 1--3~km~s$^{-1}$; Verschueren et al.\ 1997: 
0.7--1.4~km~s$^{-1}$). 
 
We also repeatedly observed two late-type stars (HD~18449, K2III,
$N=7$; HD~219615, G7III, $N=4$). The repeatability errors for these
stars are small, $0.6$ and $0.9$~km~s$^{-1}$, respectively, partly
reflecting the relative ease with which late-type spectra can be cross
correlated. The $N=7$ exposures of HD~18449 were observed on 7
different nights; the low value of the corresponding repeatability
error is consistent with the absence of significant night-to-night
instrumental zero-point shifts (Sect.~\ref{sec:data_reduction} and
Figure~\ref{fig:2}).

\begin{figure*}[t!] 
\centerline{\psfig{file=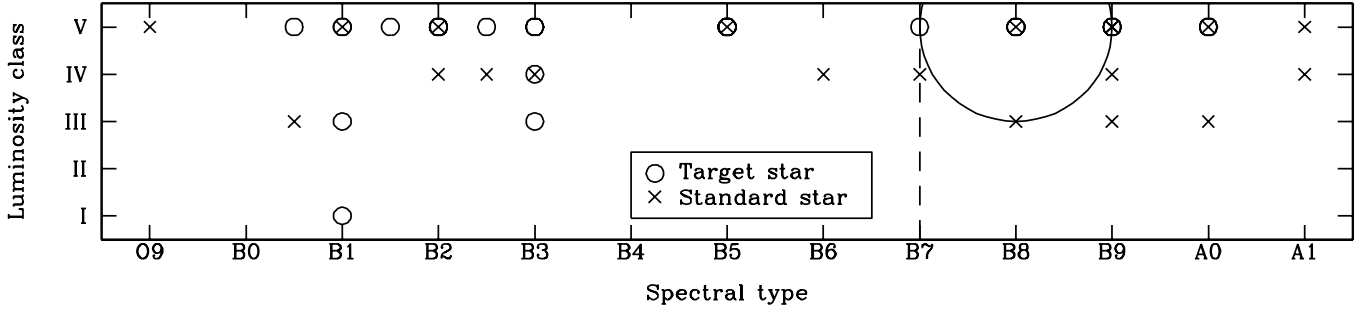,width=\textwidth,silent=}} 
\caption[]{Luminosity classes (LCl) and spectral types (SpT) of target
stars (open circles) and standard stars (crosses). This diagram is
used for selection of suitable standard stars for cross correlation.
For a given target star with luminosity class ${\rm LCl_{*}}$ and
spectral type ${\rm SpT}_{*}$, we consider all $N_{\rm s}$ standard
stars for which $d \leq d_{\rm max}$ (eq.~\ref{eq:d}). For spectral
types later than B7 ($>$B7; dashed vertical line), we use $d_{\rm max}
= 1.0$ while for stars with ${\rm SpT}_{*} = {\rm B7}$ or earlier, we
use $d_{\rm max} = 1.5$ because of sparsity of (observed) standard
stars. The circle demonstrates the selection of standard stars for a
target star with ${\rm SpT}_{*} = {\rm B8}$ and ${\rm LCl_{*}} = {\rm
V}$; for this object, we consider $N_{\rm s} = 3$ velocity standards
as appropriate templates (the three crosses falling within the circle
of radius 1). The final radial velocity $v_{\rm rad}$ for this target
star (eq.~\ref{eq:v_rad}; Table~\ref{tab:3}) is obtained by means of a
distance-weighted average of the individual radial velocities $v_{{\rm
rad},i}$ obtained using all $N_{\rm e}$ exposures of the $i =
1,\ldots,N_{\rm s}=3$ standard stars
(Sect.~\ref{subsec:mismatch_ststsel}).}
\label{fig:5} 
\end{figure*}

\renewcommand{\tabcolsep}{2.0pt} 
\begin{table*}[t] 
\caption[]{Radial velocities for 29 target stars. Each star occupies
$N$ lines, with $N$ the number of spectra/exposures obtained
(Table~\ref{tab:1}). Columns:
(1) HD number; 
(2) spectral type and luminosity class; 
(3) multiplicity (SB: spectroscopic binary; C: Hipparcos component 
    binary; G: Hipparcos acceleration binary; S: Hipparcos suspected 
    non-single --- field H61); 
(4) the number of suitable standard stars $N_{\rm s}$ and, between 
    brackets, the associated total number of exposures $N_{\rm e}$
    (Sect.~\ref{subsec:mismatch_ststsel}); 
(5) the average distance to the $N_{\rm s}$ standard stars
    (Sect.~\ref{subsec:results}); 
(6) exposure number ($1,\ldots,N$; asterisks denote suspect 
    exposures related to detector instability --- see
    Sect.~\ref{sec:data_reduction} for details); 
(7) heliocentric Julian date of the mid-point of the exposure (HJD 
    24507XX.XXXXX); 
(8) final distance-weighted radial velocity $v_{\rm rad}$ 
    (km~s$^{-1}$; eq.~\ref{eq:v_rad} in Sect.~\ref{subsec:mismatch_ststsel}); 
(9) corresponding standard deviation $\sigma_{v,{\rm rad}}$ 
    (km~s$^{-1}$; Sect.~\ref{subsec:results}); 
(10) time-averaged radial velocity $\overline{v}_{\rm rad}$ using the 
     $N$ exposures (km~s$^{-1}$; spectroscopic binaries show a dash;
     Sect.~\ref{subsec:results}); 
(11) corresponding standard deviation $\sigma_{\overline{v},{\rm 
     rad}}$ (km~s$^{-1}$; values smaller than $2.0$~km~s$^{-1}$ are 
     optimistic; Sect.~\ref{subsec:results}); 
(12) literature radial velocity (km~s$^{-1}$); 
(13) source of literature radial velocity (column~12) and remarks 
     (B52: Blaauw 1952; BvA: Blaauw \& van Albada 1963; Z83: Zentelis 
     1983).}
\begin{center} 
\begin{tabular}{rllcccrrrrrrl} 
\hline\hline 
\multicolumn{1}{c}{HD} & 
\multicolumn{1}{c}{SpT} & 
\multicolumn{1}{c}{Mult.} & 
\multicolumn{1}{c}{$N_{\rm s}$($N_{\rm e}$)} & 
\multicolumn{1}{c}{$\overline{d}$} & 
\multicolumn{1}{c}{Expos.} & 
\multicolumn{1}{c}{HJD} & 
\multicolumn{1}{c}{$v_{\rm rad}$} & 
\multicolumn{1}{c}{$\sigma_{v,{\rm rad}}$} & 
\multicolumn{1}{c}{$\overline{v}_{\rm rad}$} & 
\multicolumn{1}{c}{$\sigma_{\overline{v},{\rm rad}}$} & 
\multicolumn{1}{c}{$v_{\rm lit}$} & 
\multicolumn{1}{l}{Remark}\\ 
\multicolumn{1}{c}{(1)} & 
\multicolumn{1}{c}{(2)} & 
\multicolumn{1}{c}{(3)} & 
\multicolumn{1}{c}{(4)} & 
\multicolumn{1}{c}{(5)} & 
\multicolumn{1}{c}{(6)} & 
\multicolumn{1}{c}{(7)} & 
\multicolumn{1}{c}{(8)} & 
\multicolumn{1}{c}{(9)} & 
\multicolumn{1}{c}{(10)} & 
\multicolumn{1}{c}{(11)} & 
\multicolumn{1}{c}{(12)} & 
\multicolumn{1}{l}{(13)}\\ 
\hline 
18830  & A0      & C    & 5(16) & 0.69 & 1\phantom{*} & 61.34014 &   5.1 & 3.2 &   3.8 &   1.5 & $10.5 \pm 5.6$ & Grenier et al.\ (1999)\\ 
       &         &      & 5(16) & 0.69 & 2\phantom{*} & 61.36894 &   2.2 & 3.4 &       &        &                &\\ 
       &         &      & 5(16) & 0.69 & 3\phantom{*} & 61.39773 &   4.1 & 3.3 &       &        &                &\\ 
19216  & B9V     &      & 6(16) & 0.69 & 1\phantom{*} & 66.49641 &   7.8 & 3.4 &   8.9 &   1.3  & $ 4.5 \pm 2.7$ & Grenier et al.\ (1999)\\ 
       &         &      & 6(16) & 0.69 & 2\phantom{*} & 66.52524 &  10.3 & 3.5 &       &        &                &\\ 
       &         &      & 6(16) & 0.69 & 3\phantom{*} & 66.55355 &   8.7 & 3.6 &       &        &                &\\ 
19567  & B9      &      & 6(16) & 0.69 & 1\phantom{*} & 54.62964 &   1.7 & 3.6 &   2.6 &   1.3  &                &\\ 
       &         &      & 6(16) & 0.69 & 2*           & 54.65873 &   3.5 & 3.8 &       &        &                &\\ 
20113  & B8      &      & 4( 9) & 0.67 & 1\phantom{*} & 66.58191 &   6.1 & 4.7 &   6.2 &   2.4  &                &\\ 
       &         &      & 4( 9) & 0.67 & 2\phantom{*} & 66.61018 &   4.3 & 4.8 &        &       &                &\\ 
       &         &      & 4( 9) & 0.67 & 3*           & 66.63902 &   4.7 & 5.2 &        &       &                &\\ 
       &         &      & 4( 9) & 0.67 & 4\phantom{*} & 66.66758 &   9.5 & 4.3 &        &       &                &\\ 
20987  & B2V     & G    & 3( 7) & 0.34 & 1\phantom{*} & 54.68775 &$-22.6$& 5.7 &$-22.1$ &   0.8 &                & SpT from Abt (1985)\\ 
       &         &      & 3( 7) & 0.34 & 2\phantom{*} & 54.71652 &$-21.5$& 4.5 &        &       &                &\\ 
21483  & B3III   &      & 4( 8) & 1.05 & 1\phantom{*} & 66.36154 & $-4.0$& 3.5 & $-3.6$&   0.5  & $-6.0 \pm 0.5$ & B52\\ 
       &         &      & 4( 8) & 1.05 & 2*           & 66.39006 & $-3.2$& 3.5 &        &       &                &\\ 
21856  & B1V     &      & 4(10) & 0.86 & 1\phantom{*} & 65.36623 &  29.7 & 3.3 &  29.7  &   3.3 & $31.5 \pm 3.0$ & Z83\\ 
22114  & B8Vp    &      & 4( 9) & 0.67 & 1\phantom{*} & 65.67265 &   4.7 & 3.2 &   6.1  &   1.9 &                &\\ 
       &         &      & 4( 9) & 0.67 & 2*           & 65.70122 &   7.4 & 3.4 &        &       &                &\\ 
22951  & B0.5V   & S    & 4( 9) & 1.11 & 1\phantom{*} & 62.38453 &  19.3 & 6.3 &  19.3  &   6.3 & $19.6 \pm 3.0$ & Z83\\ 
23060  & B2Vp    &      & 5(10) & 0.55 & 1\phantom{*} & 65.41966 &  19.2 & 4.0 &  21.1  &   2.6 & $28.5 \pm 0.9$ & BvA\\ 
       &         &      & 5(10) & 0.55 & 2\phantom{*} & 65.44917 &  22.9 & 4.0 &        &       &                &\\ 
23180  & B1III   & C+SB & 4(10) & 1.00 & 1*           & 63.61336 & $-6.1$& 3.3 &     -- &     --& $12.2 \pm 0.5$ & Stickland \& Lloyd (1998)\\ 
       &         &      & 4(10) & 1.00 & 2\phantom{*} & 63.65607 &   6.4 & 3.1 &        &       &                &\\ 
       &         &      & 4(10) & 1.00 & 3\phantom{*} & 65.34880 &  80.2 & 4.2 &        &       &                &\\ 
23268  & A0      &      & 5(16) & 0.69 & 1\phantom{*} & 55.42518 &   3.8 & 3.6 &   4.1  &   0.6 &$-20.0 \pm 8.5$ & Duflot et al.\ (1995)\\ 
       &         &      & 5(16) & 0.69 & 2\phantom{*} & 55.45426 &   3.7 & 3.6 &        &       &                &\\ 
       &         &      & 5(16) & 0.69 & 3\phantom{*} & 55.48318 &   4.8 & 3.6 &        &       &                &\\ 
\hline\hline 
\end{tabular} 
\end{center} 
\label{tab:3} 
\end{table*}

\addtocounter{table}{-1} 
 
\renewcommand{\tabcolsep}{2.0pt} 
\begin{table*}[th!] 
\caption[]{Continued.} 
\begin{center} 
\begin{tabular}{rllcccrrrrrrl} 
\hline\hline 
\multicolumn{1}{c}{HD} & 
\multicolumn{1}{c}{SpT} & 
\multicolumn{1}{c}{Mult.} & 
\multicolumn{1}{c}{$N_{\rm s}$($N_{\rm e}$)} & 
\multicolumn{1}{c}{$\overline{d}$} & 
\multicolumn{1}{c}{Expos.} & 
\multicolumn{1}{c}{HJD} & 
\multicolumn{1}{c}{$v_{\rm rad}$} & 
\multicolumn{1}{c}{$\sigma_{v,{\rm rad}}$} & 
\multicolumn{1}{c}{$\overline{v}_{\rm rad}$} & 
\multicolumn{1}{c}{{$\sigma_{\overline{v},{\rm rad}}$}} & 
\multicolumn{1}{c}{$v_{\rm lit}$} & 
\multicolumn{1}{l}{Remark}\\ 
\multicolumn{1}{c}{(1)} & 
\multicolumn{1}{c}{(2)} & 
\multicolumn{1}{c}{(3)} & 
\multicolumn{1}{c}{(4)} & 
\multicolumn{1}{c}{(5)} & 
\multicolumn{1}{c}{(6)} & 
\multicolumn{1}{c}{(7)} & 
\multicolumn{1}{c}{(8)} & 
\multicolumn{1}{c}{(9)} & 
\multicolumn{1}{c}{(10)} & 
\multicolumn{1}{c}{(11)} & 
\multicolumn{1}{c}{(12)} & 
\multicolumn{1}{l}{(13)}\\ 
\hline 
23478  & B3IV... &      & 4( 8) & 0.79 & 1\phantom{*} & 65.39109 &  15.8 & 4.0 &  15.8 &   4.0 & $25.1 \pm 3.0$ & Z83; BvA: $16.4 \pm 1.1$\\ 
23597  & B8      &      & 4( 9) & 0.67 & 1\phantom{*} & 65.55700 &  17.0 & 4.9 &  16.0 &   2.6 &                &\\ 
       &         &      & 4( 9) & 0.67 & 2*           & 65.58543 &  13.0 & 4.9 &       &       &                &\\ 
       &         &      & 4( 9) & 0.67 & 3\phantom{*} & 65.61373 &  17.9 & 4.8 &       &       &                &\\ 
23625  & B2.5V   & C+SB & 5(10) & 0.76 & 1\phantom{*} & 62.40834 &   1.7 & 3.0 &    -- &    -- & $20.0 \pm 1.0$ & Blaauw \& van Hoof (1963)\\
23802  & B5Vn    & C    & 1( 2) & 0.00 & 1\phantom{*} & 54.35920 &$-53.8$& 1.8 &$-52.3$&   6.2 &                & SpT from Guetter (1977)\\ 
       &         &      & 1( 2) & 0.00 & 2\phantom{*} & 54.38800 &$-57.6$& 2.6 &       &       &                &\\ 
       &         &      & 1( 2) & 0.00 & 3\phantom{*} & 54.41664 &$-45.5$& 0.1 &       &       &                &\\ 
24012  & B5      &      & 2( 3) & 0.37 & 1\phantom{*} & 54.44633 &  26.4 & 1.0 &  26.8 &   0.5 & $36.9 \pm 1.7$ & BvA; SpT from BvA\\ 
       &         &      & 2( 3) & 0.37 & 2\phantom{*} & 54.47538 &  27.1 & 1.1 &       &       &                &\\ 
24131  & B1V     &      & 4(10) & 0.86 & 1*           & 63.57544 &  26.6 & 3.8 &  25.8 &   1.2 & $23.2 \pm 2.7$ & Z83\\ 
       &         &      & 4(10) & 0.86 & 2*           & 63.59810 &  24.9 & 3.6 &       &       &                &\\ 
24190  & B2V     & SB   & 5(10) & 0.55 & 1\phantom{*} & 62.56051 &  21.3 & 4.6 &    -- &    -- & $26.7 \pm 5.8$ & Lucy \& Sweeney (1971)\\ 
       &         &      & 5(10) & 0.55 & 2*           & 62.58902 &  22.2 & 4.6 &       &       &                &\\ 
24398  & B1Ib    &      & 1( 3) & 1.12 & 1\phantom{*} & 62.37038 &  20.1 & 1.2 &  20.1 &   1.2 & $21.6 \pm 4.1$ & Z83\\ 
24583  & B7V     &      & 4( 7) & 1.06 & 1\phantom{*} & 55.55760 &  25.7 & 4.7 &  26.2 &   5.7 & $25.4 \pm 1.4$ & BvA; SpT from Guetter (1977)\\ 
       &         &      & 4( 7) & 1.06 & 2\phantom{*} & 55.58646 &  26.7 & 4.8 &       &       &                &\\ 
       &         &      & 4( 7) & 1.06 & 3\phantom{*} & 55.61537 &  19.9 & 4.5 &       &       &                &\\ 
       &         &      & 4( 7) & 1.06 & 4\phantom{*} & 55.64456 &  35.2 & 4.8 &       &       &                &\\ 
       &         &      & 4( 7) & 1.06 & 5\phantom{*} & 55.67345 &  23.6 & 5.0 &       &       &                &\\ 
24640  & B1.5V   & S    & 5(12) & 0.87 & 1\phantom{*} & 63.45678 &  22.9 & 4.1 &  22.9 &   4.1 & $17.7 \pm 0.7$ & BvA\\ 
24970  & A0      &      & 5(16) & 0.69 & 1*           & 55.33716 &  25.3 & 3.3 &  23.2 &   2.2 & $20.4 \pm 0.3$ & Z83\\ 
       &         &      & 5(16) & 0.69 & 2\phantom{*} & 55.36676 &  23.2 & 3.5 &       &       &                &\\ 
       &         &      & 5(16) & 0.69 & 3\phantom{*} & 55.39538 &  21.0 & 3.5 &       &       &                &\\ 
25539  & B3V     &      & 4( 8) & 0.89 & 1\phantom{*} & 62.43707 &  19.0 & 3.6 &  19.0 &   3.6 & $23.8 \pm 0.3$ & Z83\\ 
25799  & B3V...  & SB   & 4( 8) & 0.89 & 1\phantom{*} & 62.48414 &  38.3 & 4.0 &    -- &    -- & $24.3 \pm 0.8$ & Morris et al.\ (1988)\\ 
       &         &      & 4( 8) & 0.89 & 2\phantom{*} & 62.51369 &  38.4 & 3.9 &       &       &                &\\ 
25833  & B5V:p   & C+SB & 2( 3) & 0.37 & 1*           & 63.53867 &  31.6 & 1.2 &    -- &    -- & $24.7 \pm 0.9$ & Popper (1974)\\ 
26499  & B9      &      & 6(16) & 0.69 & 1\phantom{*} & 54.57089 &  21.8 & 3.6 &  20.3 &   2.2 &                &\\ 
       &         &      & 6(16) & 0.69 & 2\phantom{*} & 54.60061 &  18.7 & 3.6 &       &       &                &\\ 
278942 & B3III   & C    & 2( 3) & 0.37 & 1\phantom{*} & 66.43613 &  30.8 & 2.5 &  31.4 &   0.8 &                & SpT from Cernis (1993)\\
       &         &      & 2( 3) & 0.37 & 2\phantom{*} & 66.46811 &  32.0 & 2.8 &       &       &                &\\ 
281159 & B5V     & C    & 2( 3) & 0.37 & 1\phantom{*} & 65.49889 &   9.1 & 1.8 &   8.5 &   0.9 &                & $v_{\rm rad}$ variable\\ 
       &         &      & 2( 3) & 0.37 & 2\phantom{*} & 65.52757 &   7.8 & 2.1 &       &       &                &\\ 
\hline\hline 
\end{tabular} 
\end{center} 
\end{table*}

\subsection{Template (mis)match: standard star selection} 
\label{subsec:mismatch_ststsel} 
 
A radial velocity determination by means of cross correlation of an
observed spectrum with that of a velocity standard can be biased if
the spectrum of the target does not have exactly the same
characteristics as that of the standard star. This template mismatch
can, e.g., be due to differing rotational velocities
(Sect.~\ref{subsec:mismatch_rot}). It can also result from spectral
type and/or luminosity class differences, which may translate into
differences in atmospheric velocity fields (convection, stellar wind,
pulsation, $\ldots$), in line blending, line blanketing, line
asymmetries (due to Stark broadening), or in stellar continuum.
 
We decided to use Hipparcos spectral classifications (catalogue fields
H76 and H77) throughout this study. These values are extracted from
ground-based compilations, mainly SIMBAD and the Michigan Spectral
Survey (see ESA 1997 for details), and do not necessarily have a high
and/or homogeneous quality. Classification errors (e.g., at subclass
level) can thus not be excluded. In case the Hipparcos spectral
classification lacked a luminosity class, we assumed the star was a
dwarf. We ignored classification details/peculiarities such as `p',
`n', and `e'. Comparing the Hipparcos spectral types and
luminosity classes with recent SIMBAD values and/or other literature
values did not reveal significant differences, except for the
following target stars:
HD~20987  (Hipparcos: B9 $\Rightarrow$ Abt 1985: B2V), 
HD~23802  (Hipparcos: B9 $\Rightarrow$ Guetter 1977: B5Vn), 
HD~24012  (Hipparcos: B9 $\Rightarrow$ Blaauw \& van Albada 1963: B5), 
HD~24583  (Hipparcos: A0 $\Rightarrow$ Guetter 1977: B7V), and 
HD~278942 (Hipparcos: F2 $\Rightarrow$ Cernis 1993: B3III).
We discuss these stars individually in Sect.~\ref{sec:special}.
We decided, for these five `problem cases', to use the spectral
classifications derived in the dedicated studies listed
above. Alternatively, we could have used, e.g., the strength of the
helium lines in our spectra as a coarse spectral type indicator (for
example, the ratio He{\small{\sc{I}}} 4471/Mg{\small{\sc{II}}} 4481,
not accessible from our data, is often used for determining the
spectral types of stars; Jaschek \& Jaschek 1990). In case of line 
anomalies, line indicators may provide misleading classifications. 
B-type stars, and notably the helium lines in
their spectra, are known to be particularly sensitive to these
problems. Dozens of peculiar stars, for example in the form of
helium-strong and helium-weak stars, have been discovered in the
nearby OB associations (Jaschek \& Jaschek 1990), and we therefore
decided not to use helium line strengths as spectral type indicators.

In order to investigate the effect of spectral mismatch (including
luminosity class mismatch), we performed a cross correlation of all
possible combinations of all exposures of all 20 standard stars. After
averaging over exposure pairs, we constructed a $20 \times 20$
`matrix' of average shifts between all pairs of standard stars. The
diagonal elements in this matrix, which is skew-symmetric, correspond
to the repeatability errors (Sect.~\ref{subsec:reper}). After ordering
the stars on spectral type, one can see that shifts grow when moving
away from the diagonal. However, close to the diagonal (i.e., within a
few matrix elements, i.e., generally within a few sub-classes), shifts
are mostly relatively small (they are even generally insignificant in
this exercise as a result of random errors). The results of this
analysis indicate, nonetheless, that spectral mismatch can easily give
rise to 5--10~km~s$^{-1}$ systematic errors or larger. For this
reason, we decided, for a given target star exposure, not to use a
single radial velocity standard as template. Instead, we used a set of
standard stars with spectral types and luminosity classes `similar to'
the target, and averaged the corresponding radial velocities to one,
final value. This has the additional advantage that radial velocity
zero-point differences/errors in our standard star sample are washed
out to some degree.
 
In order to quantify the meaning of `similar spectral types and 
luminosity classes', we defined a `distance' $d$ between any standard 
and any target star: 
\begin{equation} 
d \equiv \sqrt{({\rm SpT_{*}}-{\rm SpT})^2 + {{1}\over{4}} \cdot ({\rm 
LCl_{*}}-{\rm LCl})^2}. 
\label{eq:d} 
\end{equation} 
We then constructed a luminosity class--spectral type (LCl--SpT)
diagram containing all target and standard stars (Figure~\ref{fig:5}).
From this diagram, we decided to define all standard stars for which
$d \leq d_{\rm max}$ as {\it suitable templates} for a target star
with luminosity class ${\rm LCl_{*}}$ and spectral type ${\rm
SpT}_{*}$. The unit of the distance $d$ was chosen such that a
difference of one spectral type subclass and a difference of two
luminosity classes both denote a distance of 1 (e.g., `${\rm B8}-{\rm
B6} \rightarrow d = 2$' and `${\rm V}-{\rm III} \rightarrow d =
1$'). The maximum distance ($d_{\rm max}$) was chosen as small as
possible, but still large enough for most target stars to have more
than one associated standard star:
\begin{equation} 
d_{\rm max} = \left\{
\begin{array}{cl} 
1.5 & \qquad {\rm for~SpT}_{*} \leq {\rm B7},\\ 
1.0 & \qquad {\rm for~SpT}_{*} > {\rm B7},\\ 
\end{array} 
\right.
\end{equation} 
where $\leq$${\rm B7}$ means B7 or earlier and $>$${\rm B7}$ means
later than B7. The radial velocity $v_{\rm rad}$ for the target is
obtained by means of a distance-weighted average of the individual
radial velocities $v_{{\rm rad},i}$ obtained by cross correlating with
all $N_{\rm e}$ exposures of the $i = 1, \ldots, N_{\rm s}$ suitable
standard stars:
\begin{equation} 
v_{\rm rad} = {{\sum_{i = 1}^{N_{\rm e}} w_i \cdot
               v_{{\rm rad},i}}\over{\sum_{i = 1}^{N_{\rm e}} w_i}}, 
\label{eq:v_rad} 
\end{equation} 
where 
\begin{equation} 
w_i = d_{\rm max} - 0.5 \cdot d_i.
\label{eq:w} 
\end{equation} 
This choice for $w_i$ has the property that $w_i (d_i = 0) / w_i (d_i 
= d_{\rm max}) = 2$, independent of spectral type. 

As an example, Figure~\ref{fig:3} shows the spectra of HD~1438
and HD~24583; the first star (B8V) has been used as a standard in the
radial velocity determination of the latter (B7V). Although the
signal-to-noise ratios and rotational velocities of the two
spectra/stars differ, minor spectral differences are clearly visible,
e.g., in the helium line at 4026~\AA. The `spectral-type-distance' $d$
between the stars equals $1$ (eq.~\ref{eq:d}), which is the maximum
allowed distance ($d_{\rm max}$) for targets with spectral types B7V
and later.

\subsection{Template (mis)match: stellar rotation} 
\label{subsec:mismatch_rot} 
 
We carried out a simple experiment to investigate the effect of
rotational mismatch on the final radial velocities. We selected the
H$\delta$ absorption line in the spectrum of the slow rotator HD~38899
(B9IV; $v \sin i = 15$--30~km~s$^{-1}$). We broadened this line, so as
to mimic an increasing rotation velocity, with a Gaussian with
increasing full-width at half maximum. We then cross correlated the
broadened line with the original line and found no significant
velocity shifts (compared to the expected repeatability error of
$\overline{\sigma_{\rm r}} = 2.1$~km~s$^{-1}$), even for a broadening
parameter which corresponded to a rotation velocity exceeding the
break-up velocity. We repeated this exercise for HD~35708 (B2.5IV; $v
\sin i = 10$--24~km~s$^{-1}$) with similar results. Although this
experiment is simplistic (it neglects, e.g., noise and effects due to
changes in the limb darkening), we conclude that rotational mismatch
is not relevant at the level of 1--2~km~s$^{-1}$ (see also Verschueren
\& David 1999). Hence, our `bias' in preferentially selecting slowly
rotating standards (Sect.~\ref{sec:obs}) is not significant.
 
\subsection{Results} 
\label{subsec:results} 
 
Table~\ref{tab:3} lists the results of applying the procedure
described in Sect.~\ref{subsec:mismatch_ststsel} to the available
target spectra. Table~\ref{tab:3} also lists $N_{\rm s}$ and $N_{\rm
e}$, the number of standard stars and the number of standard star
exposures used in determining $v_{\rm rad}$, respectively, and the
mean radial velocity for each exposure (eq.~\ref{eq:v_rad}); the
corresponding standard deviation is denoted $\sigma_{v,{\rm
rad}}$. The mean distance in luminosity class--spectral type space
(Figure~\ref{fig:5}) between the target and the $N_{\rm e}$ standard
star exposures is denoted $\overline{d} \equiv {N_{\rm e}}^{-1} \cdot
\sum_{i = 1}^{N_{\rm e}} d_i$. In general, the effect of template
mismatch increases with increasing $\overline{d}$. Table~\ref{tab:3}
also lists, for each target star, the time-averaged radial velocity
$\overline{v}_{\rm rad}$, and the corresponding standard deviation
$\sigma_{\overline{v},{\rm rad}}$, of the $N$ radial velocities
corresponding to the $N$ exposures obtained for each star. The
quantity $\overline{v}_{\rm rad}$ is the best estimate of the true
radial velocity of an object, provided it is single;
$\overline{v}_{\rm rad}$ is potentially less straightforward to
interpret otherwise.
 
Table~\ref{tab:3} shows that, in general, $\sigma_{v,{\rm rad}} \gg
\sigma_{\overline{v},{\rm rad}} \ga \overline{\sigma_{\rm r}}$. The
first inequality suggests that template mismatch, combined with
possible errors in the assumed radial velocities for the standard
stars, is a significant error source, justifying our
averaging-approach. The similar magnitudes of
$\sigma_{\overline{v},{\rm rad}}$ (with a mean value for all stars of
1.5~km~s$^{-1}$) and the mean repeatability error
($\overline{\sigma_{\rm r}} = 2.1$~km~s$^{-1}$) provide a further
indication of the validity of our approach.
 
The errors $\sigma_{\overline{v},{\rm rad}}$ on $\overline{v}_{\rm
rad}$ are sometimes remarkably small, down to a few tenths of a
km~s$^{-1}$. Given that template mismatch can have an effect at the
level of $\sim$$3.0$--$5.0$~km~s$^{-1}$ (corresponding to
$\sigma_{v,{\rm rad}}$), that the mean repeatability error is
$2.1$~km~s$^{-1}$ (Sect.~\ref{subsec:reper}), and that the expected
error due to uncertainties in the radial velocities of the standard
stars is also $\sim$$2.0$~km~s$^{-1}$
(Sect.~\ref{subsec:standard_stars}), we suspect that increasing
$\sigma_{\overline{v},{\rm rad}}$ to $2.0$~km~s$^{-1}$ provides a more
realistic estimate of the error.

The 11 target exposures identified in Sect.~\ref{sec:data_reduction}
as suspect as a result of detector instability are indicated with
asterisks in column~6 of Table~\ref{tab:3}. Comparing the
associated radial velocities with non-suspect exposures shows no
significant deviations, consistent with the relatively small magnitude
of the effect established in Sect.~\ref{sec:data_reduction}
($0.6$--$2.9$~km~s$^{-1}$; Figure~\ref{fig:2}). We therefore do not
discriminate these suspect exposures further and treat them as normal
in the remainder of this manuscript. They thus also contribute to the
time-averaged radial velocities $\overline{v}_{\rm rad}$.
 
\renewcommand{\tabcolsep}{15.00pt} 
\begin{table*}[th!] 
\caption[]{Illustration of results underlying the mean radial
velocities $v_{\rm rad}$ in Table~\ref{tab:3} (cf.\
eq.~\ref{eq:v_rad}). The star HD~24970 (A0) was observed $N = 3$ times
and has $N_{\rm s} = 5$ suitable standard stars, giving rise to 5
radial velocities per exposure. These values are listed (in km~s$^{-1}$),
together with the spectral classifications and distances $d$ and
weights $w$ (see eqs~\ref{eq:d} and \ref{eq:w}) of the standard
stars. The distance-weighted mean radial velocities with their errors
are listed in the final column (in km~s$^{-1}$). Systematic
differences between columns in this table can be due to template
mismatch and/or incorrect literature radial velocities; in general,
systematic differences between lines in this table can be due to
random errors (statistics), instrumental shifts, and/or multiplicity
(not likely in this specific case).}
\begin{center} 
\begin{tabular}{crrrrrr} 
\hline\hline 
\multicolumn{1}{r}{HD} & 
\multicolumn{1}{c}{10982} & 
\multicolumn{1}{c}{196724} & 
\multicolumn{1}{c}{196821} & 
\multicolumn{1}{c}{27638} & 
\multicolumn{1}{c}{58142} & 
\multicolumn{1}{c}{$v_{\rm rad}$}\\ 
\multicolumn{1}{r}{SpT} & 
\multicolumn{1}{c}{B9V} & 
\multicolumn{1}{c}{A0V} & 
\multicolumn{1}{c}{A0III} & 
\multicolumn{1}{c}{B9V} & 
\multicolumn{1}{c}{A1V} & 
\multicolumn{1}{c}{}\\ 
\multicolumn{1}{r}{$d=$} & 
\multicolumn{1}{c}{$1.00$} & 
\multicolumn{1}{c}{$0.00$} & 
\multicolumn{1}{c}{$1.00$} & 
\multicolumn{1}{c}{$1.00$} & 
\multicolumn{1}{c}{$1.00$} & 
\multicolumn{1}{c}{}\\ 
\multicolumn{1}{r}{$w=$} & 
\multicolumn{1}{c}{$0.50$} & 
\multicolumn{1}{c}{$1.00$} & 
\multicolumn{1}{c}{$0.50$} & 
\multicolumn{1}{c}{$0.50$} & 
\multicolumn{1}{c}{$0.50$} & 
\multicolumn{1}{c}{}\\ 
\hline 
{\rm Exposure\ } 1*           & 26.2~ & 27.8~ & 24.3~ & 21.4~ & 
22.8~ & $25.3 \pm 3.3$\\ 
{\rm Exposure\ } 2\phantom{*} & 24.1~ & 26.1~ & 21.6~ & 20.1~ & 
20.2~ & $23.2 \pm 3.5$\\ 
{\rm Exposure\ } 3\phantom{*} & 21.6~ & 23.8~ & 19.6~ & 16.9~ & 
18.4~ & $21.0 \pm 3.5$\\ 
\hline\hline 
\end{tabular} 
\end{center} 
\label{tab:4} 
\end{table*}
 
\renewcommand{\tabcolsep}{15.00pt} 
\begin{table*}[th!] 
\caption[]{The standard deviations of the radial velocities
corresponding to the $N_{\rm e}$ different exposures of the $N_{\rm s}
= 5$ suitable standard stars for all combinations of the different
object exposures and suitable standard stars of the stars HD~24970
(A0; cf.\ Table~\ref{tab:4}; top part) and HD~23268 (A0; bottom
part). The distance-weighted mean radial velocities with their errors
are listed in the final column (in km~s$^{-1}$); time-averaged values,
denoted $\overline{v}_{\rm rad}$, are provided in
Table~\ref{tab:3}. The quantity $\sigma_{\rm r}$ denotes the
repeatability error (Sect.~\ref{subsec:reper}).}
\begin{center} 
\begin{tabular}{ccccccr} 
\hline\hline 
\multicolumn{1}{r}{HD} & 
\multicolumn{1}{c}{10982} & 
\multicolumn{1}{c}{196724} & 
\multicolumn{1}{c}{196821} & 
\multicolumn{1}{c}{27638} & 
\multicolumn{1}{c}{58142} & 
\multicolumn{1}{c}{$v_{\rm rad}$}\\ 
\multicolumn{1}{r}{SpT} & 
\multicolumn{1}{c}{B9V} & 
\multicolumn{1}{c}{A0V} & 
\multicolumn{1}{c}{A0III} & 
\multicolumn{1}{c}{B9V} & 
\multicolumn{1}{c}{A1V} & 
\multicolumn{1}{c}{}\\ 
\multicolumn{1}{r}{$d=$} & 
\multicolumn{1}{c}{$1.00$} & 
\multicolumn{1}{c}{$0.00$} & 
\multicolumn{1}{c}{$1.00$} & 
\multicolumn{1}{c}{$1.00$} & 
\multicolumn{1}{c}{$1.00$} & 
\multicolumn{1}{c}{}\\ 
\multicolumn{1}{r}{$w=$} & 
\multicolumn{1}{c}{$0.50$} & 
\multicolumn{1}{c}{$1.00$} & 
\multicolumn{1}{c}{$0.50$} & 
\multicolumn{1}{c}{$0.50$} & 
\multicolumn{1}{c}{$0.50$} & 
\multicolumn{1}{c}{}\\ 
\multicolumn{1}{r}{$N=$} & 
\multicolumn{1}{c}{$2$} & 
\multicolumn{1}{c}{$5$} & 
\multicolumn{1}{c}{$3$} & 
\multicolumn{1}{c}{$2$} & 
\multicolumn{1}{c}{$4$} & 
\multicolumn{1}{c}{}\\ 
\multicolumn{1}{r}{$\sigma_{\rm r}=$} & 
\multicolumn{1}{c}{$3.4$} & 
\multicolumn{1}{c}{$2.3$} & 
\multicolumn{1}{c}{$0.9$} & 
\multicolumn{1}{c}{$0.2$} & 
\multicolumn{1}{c}{$2.0$} & 
\multicolumn{1}{c}{}\\ 
\hline 
{\rm Exposure\ } 1*           & 1.8 & 3.4 & 0.4 & 0.2 & 2.7 & $+25.3 \pm 3.3$\\ 
{\rm Exposure\ } 2\phantom{*} & 2.3 & 3.5 & 0.9 & 0.6 & 2.7 & $+23.2 \pm 3.5$\\ 
{\rm Exposure\ } 3\phantom{*} & 2.3 & 3.5 & 0.9 & 0.7 & 2.6 & $+21.0 \pm 3.5$\\ 
\hline		        	    	   	  	 	
{\rm Exposure\ } 1\phantom{*} & 2.1 & 3.5 & 0.8 & 0.1 & 2.6 &  $+3.8 \pm 3.6$\\ 
{\rm Exposure\ } 2\phantom{*} & 2.2 & 3.5 & 1.0 & 0.6 & 2.6 &  $+3.7 \pm 3.6$\\ 
{\rm Exposure\ } 3\phantom{*} & 2.3 & 3.6 & 1.0 & 0.6 & 2.6 &  $+4.8 \pm 3.6$\\ 
\hline\hline 
\end{tabular} 
\end{center} 
\label{tab:5} 
\end{table*}

Table~\ref{tab:4} shows, as a representative example, how the final, 
mean radial velocities $v_{\rm rad}$ in Table~\ref{tab:3} 
(eq.~\ref{eq:v_rad}) are built up for the A0 star HD~24970. This 
star was observed $N=3$ times and has $N_{\rm s} = 5$ suitable 
standard stars (Figure~\ref{fig:5}), giving rise to $N_{\rm s} = 5$ 
radial velocities per exposure. The best-match radial velocity standard 
(HD~196724, A0V, $d=0.00$, and $w=1.00$) implies the target star's 
radial velocity is $\sim$$26$~km~s$^{-1}$. The other four standards, 
however, which have rather closely matching spectral classifications, 
imply a velocity of $\sim$$20$--$25$~km~s$^{-1}$. Two explanations, 
which are not mutually exclusive, may be invoked to 
reconcile this discrepancy: either template mismatch explains the 
systematically lower values, or an error in the literature radial 
velocity of the A0V standard HD~196724 explains the systematically 
higher value. Discriminating between these two options is impossible 
using the available data. The problem is, however, automatically 
`solved' for/alleviated in the weighted-average approach adopted 
here. 
 
Table~\ref{tab:5} shows, for the single A0 stars HD~24970 ($N=3$; 
$\sigma_{\rm r} = 2.7$~km~s$^{-1}$) and 23268 ($N=3$; $\sigma_{\rm r} 
= 1.4$~km~s$^{-1}$), for all combinations of the $N = 3$ object 
exposures and the $N_{\rm s} = 5$ suitable standard stars, the standard 
deviations of the radial velocities corresponding to the different 
exposures of the {\it standard\/} stars (cf.\ Table~\ref{tab:4}). The 
standard deviations are, in some cases, slightly larger than the 
expected values (roughly the repeatability errors), suggesting that 
both errors in the literature radial velocities of standard stars and 
template mismatch contribute significantly to the total error 
budget. 

\renewcommand{\tabcolsep}{5.5pt} 
\begin{table*}[th!] 
\caption[]{The five spectroscopic binaries (SBs) and their orbital
elements. Columns for the first part of the table (literature data):
(1) HD number; 
(2) periastron date $T_0$ (HJD); 
(3) period $P$ (days); 
(4) primary orbital semi-amplitude $K_1$ (km~s$^{-1}$); 
(5) secondary orbital semi-amplitude $K_2$ (km~s$^{-1}$; only for
    double-lined spectroscopic binaries, i.e., SB2s); 
(6) systemic velocity $\gamma$ (km~s$^{-1}$); 
(7) orbital eccentricity $\epsilon$ (see the footnote to the table for
    literature sources). 
Columns for the second part of the table (data from this study): 
(1) HD number; 
(2) exposure number (Table~\ref{tab:3}); 
(3) instantaneous radial velocity (km~s$^{-1}$); 
(4) primary component prediction using the ephemeris provided above
    (km~s$^{-1}$);
(5) secondary component prediction using the ephemeris provided above
    (km~s$^{-1}$; only for SB2s).}
\begin{center} 
\begin{tabular}{cccclcl} 
\hline\hline 
\multicolumn{1}{c}{HD} & 
\multicolumn{1}{c}{$T_0$~[HJD]} & 
\multicolumn{1}{c}{$P$~[days]} & 
\multicolumn{1}{c}{$K_1$~[km~s$^{-1}$]} & 
\multicolumn{1}{c}{$K_2$~[km~s$^{-1}$]} & 
\multicolumn{1}{c}{$\gamma$~[km~s$^{-1}$]} & 
\multicolumn{1}{c}{$\epsilon$}\\ 
\multicolumn{1}{c}{(1)} & 
\multicolumn{1}{c}{(2)} & 
\multicolumn{1}{c}{(3)} & 
\multicolumn{1}{c}{(4)} & 
\multicolumn{1}{c}{(5)} & 
\multicolumn{1}{c}{(6)} & 
\multicolumn{1}{c}{(7)} \\ 
\hline 
23180 & $2445338.267\phantom{0} \pm 0.005$ & $ \phantom{0}4.4191447 \pm 0.0000082$ & $111.8 \pm 1.4$ & $155.0 \pm 2.4$ & $+12.2 \pm 0.5$ & $0              $\\ 
23625 & $2400000.0\phantom{000}   \pm 0.005$ & $ \phantom{0}1.940564\phantom{0}  \pm 0.000004\phantom{0} $ & $ \phantom{0}82.0 \pm 2.0$ & $114.0 \pm 3.0$ & $+20.0 \pm 1.0$ & $0              $\\ 
24190 & $2435401.00\phantom{00}  \pm 0.01\phantom{0} $ & $26.111\phantom{0000}     \pm 0.001\phantom{0000}    $ & $ \phantom{0}13.0 \pm 1.0$ & --              & $+26.7 \pm 1.0$ & $0.08\phantom{0}  \pm 0.01\phantom{0} $\\ 
25799 & $2435152.835\phantom{0} \pm 0.013$ & $ \phantom{0}0.9121679 \pm 0.0000019$ & $ \phantom{0}17.6 \pm 1.2$ & --              & $+24.3 \pm 0.8$ & $0              $\\ 
25833 & $2424946.5150 \phantom{ \pm 0.0000}        $ & $ \phantom{0}2.0287298 \phantom{ \pm 0.0000000\,\,}             $ & $162.8 \pm 1.1$ & $178.7 \pm 1.2$ & $+24.7 \pm 0.9$ & $0.071 \pm 0.001$\\
\hline\hline 
\end{tabular}
\end{center}
\vskip-0.26truecm
HD~23180: Stickland \& Lloyd (1998);\hfill\break 
HD~23625: Blaauw \& van Hoof (1963);\hfill\break 
HD~24190: Lucy \& Sweeney (1971);\hfill\break 
HD~25799: Morris et al.\ (1988); this is a non-radial pulsator;\hfill\break 
HD~25833: Gim\'enez \& Clausen 1994 (cf.\ Popper 1974).\hfill\break
\null\vskip-2.25truecm
\begin{center}
\begin{tabular}{ccrrr}
\hline\hline 
\multicolumn{1}{c}{HD} & 
\multicolumn{1}{c}{Exposure} & 
\multicolumn{1}{c}{$v_{\rm rad}$~[km~s$^{-1}$]} & 
\multicolumn{1}{c}{$v_{\rm rad,1}$~[km~s$^{-1}$]} & 
\multicolumn{1}{c}{$v_{\rm rad,2}$~[km~s$^{-1}$]} \\ 
\multicolumn{1}{c}{(1)} & 
\multicolumn{1}{c}{(2)} & 
\multicolumn{1}{c}{(3)} & 
\multicolumn{1}{c}{(4)} & 
\multicolumn{1}{c}{(5)} \\ 
\hline 
23180 & 1*           & $-6.1 \pm 3.3$ & $  \phantom{\sim 0}-28.0 \pm  \phantom{0}1.8$ & $  67.9 \pm  \phantom{0}2.5$\\ 
      & 2\phantom{*} & $ 6.4 \pm 3.1$ & $  \phantom{\sim 0}-21.6 \pm  \phantom{0}1.8$ & $  59.0 \pm  \phantom{0}2.5$\\ 
      & 3\phantom{*} & $80.2 \pm 4.2$ & $  \phantom{\sim}108.7 \pm  \phantom{0}1.6$   & $-121.6 \pm  \phantom{0}2.5$\\ 
23625 & 1\phantom{*} & $ 1.7 \pm 3.0$ & $  \phantom{\sim 0}-34.7 \pm 19.1$ & $  96.1 \pm 26.5$\\ 
24190 & 1\phantom{*} & $21.3 \pm 4.6$ & $  \phantom{\sim 0}20.6 \pm  \phantom{0}1.9$ & --                 \\ 
      & 2*           & $22.2 \pm 4.6$ & $  \phantom{\sim 0}20.5 \pm  \phantom{0}1.9$ & --                 \\ 
25799 & 1\phantom{*} & $38.3 \pm 4.0$ & $  \phantom{\sim 0}40.3 \pm  \phantom{0}2.1$ & --                 \\ 
      & 2\phantom{*} & $38.4 \pm 3.9$ & $  \phantom{\sim 0}41.2 \pm  \phantom{0}1.7$ & --                 \\ 
25833 & 1*           & $31.6 \pm 1.2$ & $\sim$$190 \phantom{ \pm 00.00\,\,\,\,\,}$ & $\sim$$-150 \phantom{ \pm 00.00\,\,\,\,\,}$\\ 
\hline\hline 
\end{tabular} 
\end{center}
\label{tab:6} 
\end{table*}

\subsection{Spectroscopic binaries} 
\label{sec:sbs} 

Our sample contains five known spectroscopic binaries (SBs; a sixth
SB, X~Per/HD~24534, was classified as Per~OB2 member by B52, but was
not observed by us; see Sect.~\ref{subsec:B52}). Three of these have
previously been identified as double-lined SBs (SB2s). The modest
resolving power of the spectrograph-grating combination used by us ($R
\sim 7000$), combined with the very broad hydrogen absorption lines in
the spectra of these SB2s, do not allow a proper decomposition of the
spectra in all of these cases (we did not optimize our instrumental
setup for SB2s). We were therefore forced to treat these spectra as
arising from a single star. The resulting radial velocities, derived
by means of cross correlation using a template star matched to the
spectral type and luminosity class of the primary component, are of
modest physical significance. Conceptually, they correspond to a
luminosity-weighted average of the instantaneous radial velocities of
the primary and secondary components, with template mismatch
complicating matters (recall that the standard star was selected based
on the primary component exclusively). While the composite spectrum we
observed is a luminosity-weighted average of the instantaneous
Doppler-shifted spectra of the two components, the inferred radial
velocity is not necessarily a luminosity-weighted average of the
instantaneous radial velocities of the components, although a trend
along these lines might be expected.
 
The orbital elements from the literature for the five known SBs in our
sample are listed in Table~\ref{tab:6}. We have one or a few
measurements for each of these objects. Repeat measurements were
usually taken on the same night, typically within one hour, with the
exception of o~Per (Table~\ref{tab:3}). As shown in
Table~\ref{tab:6}, the amplitudes of the radial velocity variations
are large, ranging up to $\sim$$180$~km~s$^{-1}$ for AG~Per. With the
caveat mentioned above for SB2s, our instantaneous measurements fall
within these ranges in all five cases.
 
For the two single-lined SBs (the SB1s), we used the known periods to 
investigate whether (i) our repeat measurements should have shown 
significant differences, and (ii) whether we could reconstruct what 
the expected $v_{\rm rad}$ should have been at the epoch of our 
observations. As the periods are known to sufficient accuracy, it is 
possible to do this, and we find agreement in both cases. We discuss 
each binary in more detail below. 
 
\smallskip\noindent{\bf HD~23180 (o~Per):\/} This well-known SB2
(e.g., Stickland \& Lloyd 1998) was observed twice (the first exposure
being suspect), and then once more, $\sim$$1.75$ days later. According
to Hipparcos, o~Per is also an $1{\farcs}019$ visual binary with
$\Delta Hp = 2.91$~mag (Table~\ref{tab:1}; see Stickland \& Lloyd for
a discussion of this component). The Hipparcos data imply that our
observations have primarily ($\sim$93\% in flux) detected photons from
the B1III + B2V SB. A simple-minded comparison of our radial
velocities, obtained by simply ignoring the presence of the secondary
component, with predictions based on Stickland \& Lloyd's (1998)
orbital ephemeris shows the expected `flux-weighted-average' trend
(Table~\ref{tab:6}). The third exposure of o~Per shows some double
helium lines with an average velocity separation of $\sim$$200 \pm
30$~km~s$^{-1}$; this is consistent with the expected velocity
separation for this exposure of $\sim$$230$~km~s$^{-1}$
(Table~\ref{tab:6}).

\smallskip\noindent{\bf HD~23625:\/} This object was studied by Blaauw
\& van Hoof (1963), who classified it as a B2V SB2. HD~23625 is also a
Hipparcos component binary with $\Delta Hp = 2.58$~mag and $\rho =
3{\farcs}349$ (Table~\ref{tab:1}); these data imply that this fainter
component cannot have been present in the $3^{\prime\prime}$-diameter
entrance pupil of the image slicer. We observed this object once, and
find $v_{\rm rad} = 1.7 \pm 3.0$~km~s$^{-1}$ (our spectrum
superficially looks single-lined, although a slight asymmetry in the
line profiles is present; the radial velocity was again obtained by
ignoring the presence of the secondary component in the
spectrum). According to Blaauw \& van Hoof's ephemeris, we should have
obtained $-34.7 \pm 19.1$~km~s$^{-1}$ for the primary
component. This prediction agrees with our measurement at the
2$\sigma$-level.
 
\smallskip\noindent{\bf HD~24190:\/} This object was identified as a
low-amplitude B2V SB1 by Blaauw \& van Albada (1963; BvA). According
to the improved ephemeris from Lucy \& Sweeney (1971), our exposures
 ($21.3 \pm 4.6$ and $22.2 \pm 4.6$~km~s$^{-1}$) should have read
$20.6 \pm 1.9$ and $20.5 \pm 1.9$~km~s$^{-1}$, respectively. We
conclude our measurements are in full agreement with Lucy \& Sweeney's
ephemeris.
 
\smallskip\noindent{\bf HD~25799:\/} This object was classified as a
low-amplitude eclipsing B3V SB1 with a poorly determined/variable
period by BvA ($P \sim 10.67$~d and $K_1 = 20$~km~s$^{-1}$). BvA's
discovery of `erratic velocity changes' was in fact the first
indication that this object is not a normal binary. Morris et al.\
(1988) showed that an improved ephemeris (notably $P = 0.9121679$~d)
could fit all previous spectroscopic data, but also showed, based on
new photometric data, that the object is not a binary but a non-radial
pulsator. This finding was confirmed later by Hipparcos photometry. We
observed this object twice, obtaining $v_{\rm rad} = 38.3 \pm
4.0$~km~s$^{-1}$ and $38.4 \pm 3.9$~km~s$^{-1}$. These observations
are consistent with Morris et al.'s ephemeris within $1\sigma$, which
predicts $v_{\rm rad} = 40.3 \pm 2.1$ and $41.2 \pm
1.7$~km~s$^{-1}$, respectively.
 
\smallskip\noindent{\bf HD~25833 (AG~Per):\/} This object is one of
the few detached massive eclipsing SB2s, and is very well-studied
(e.g., Popper 1974; cf.\ Popper \& Hill 1991). The system is eccentric
($e = 0.071 \pm 0.001$) and shows apsidal motion with a period of
$75.6 \pm 0.6$~yr (i.e., $4.76^{\circ\/}$~yr$^{-1}$). Spectroscopic
ephemeris predictions are non trivial as a result. As both components
are roughly equally bright ($V \sim 7.46$ and $\sim$$7.88$~mag;
Table~6 in Gim\'enez \& Clausen 1994) and have similar spectral types
(B4V + B5; Gim\'enez \& Clausen), we expect the radial velocity
inferred from our low-resolution composite spectrum to be near the
systemic velocity, with a relatively small `flux-weighted-average'
oscillation due to orbital motion. Additional complications might
arise from the presence of a third component, with Hipparcos
parameters $\Delta Hp = 1.81$~mag and $\rho = 0{\farcs}803$
(Table~\ref{tab:1}). Gim\'enez \& Clausen (1994) provide the most
recent photometric eclipse ephemeris, from which we derive that the
single, formally suspect, exposure we took of this object should have
an associated phase of $\sim$$0.20$ with respect to a secondary
eclipse. This roughly implies radial velocities for the primary of
$\gamma + K_1 \sim +24.7 + 162.8 \sim +190$~km~s$^{-1}$ and for the
secondary of $\gamma - K_2 \sim +24.7 - 178.7 \sim -150$~km~s$^{-1}$,
so that the expected velocity separation is
$\sim$$340$~km~s$^{-1}$. Not surprisingly, the hydrogen lines in our
composite spectrum do not show clear double-lined signs (although an
asymmetry of the line profiles seems present), but the much narrower
helium lines are double. From these, we derive a mean velocity
separation of $329 \pm 17$~km~s$^{-1}$ (based on 5 double lines),
consistent with expectations. Ignoring the presence of the secondary
component in the composite spectrum, we derive $v_{\rm rad} =
31.6 \pm 1.2$~km~s$^{-1}$ (based mainly on the broad hydrogen lines
in the spectrum), which is indeed close to the systemic velocity.

\subsection{Comments on individual stars} 
\label{sec:special} 

We noted in Sect.~\ref{subsec:mismatch_ststsel} that five targets have
Hipparcos spectral types and luminosity classes which differ
significantly from other literature sources. In all these cases, we
dropped the Hipparcos data and adopted the classifications from the
dedicated studies. The implications of this choice for the inferred
radial velocities are discussed below.
 
\smallskip\noindent{\bf HD~20987:\/} The Hipparcos spectral type (B9)
suggests $v_{\rm rad} \sim -40$~km~s$^{-1}$. Abt (1985) and Roman
(1978), independently, quote a spectral classification of B2V, which
suggests $v_{\rm rad} = -22.1 \pm 0.8$~km~s$^{-1}$
(Table~\ref{tab:3}; we used $d_{\rm max} = 0.75$).
 
\smallskip\noindent{\bf HD~23802:\/} The Hipparcos spectral type (B9)
suggests $v_{\rm rad} \sim -70$ to $-80$~km~s$^{-1}$. Guetter (1977)
quotes a spectral classification of B5Vn, which results in 
$v_{\rm rad} = -52.3 \pm 6.2$~km~s$^{-1}$ (Table~\ref{tab:3}; we used
$d_{\rm max} = 0.75$). The three exposures give a relatively large
spread in the inferred radial velocities ($\sim$$12$~km~s$^{-1}$),
which, if significant, suggests that the radial velocity of this
object is variable, most likely as a result of duplicity.
 
\smallskip\noindent{\bf HD~24012:\/} The Hipparcos spectral type (B9)
suggests $v_{\rm rad} \sim 19$~km~s$^{-1}$. BvA quote a spectral
classification of B5, which suggests $v_{\rm rad} = 26.8 \pm
0.5$~km~s$^{-1}$ (Table~\ref{tab:3}).
 
\smallskip\noindent{\bf HD~24583:\/} The Hipparcos spectral type (A0)
suggests $v_{\rm rad} \sim 22$~km~s$^{-1}$. Guetter (1977) quotes a
spectral classification of B7V, which suggests $v_{\rm rad} =
26.2 \pm 5.7$~km~s$^{-1}$ (Table~\ref{tab:3}). Our five exposures
seem to contain 1--2 outliers (exposure~4 and, to a smaller extent,
exposure~3). Excluding exposure~4 returns ${\overline v}_{\rm
rad} = 24.0 \pm 3.0$~km~s$^{-1}$; excluding both exposures returns
${\overline v}_{\rm rad} = 25.4 \pm 1.6$~km~s$^{-1}$.
 
\medskip\noindent{\bf HD~278942:\/} This faint object is a Hipparcos
component binary ($\rho = 0{\farcs}149$ and $\Delta Hp = 0.78$~mag;
Table~\ref{tab:1}). An IRAS ring in the interstellar medium around
this star explains the various colour and spectral type measurements
reported in the literature (see Z99 for details; Hipparcos/Tycho: SpT
= F2 and $B-V = 1.13$~mag; SIMBAD/AGK3: SpT = B5 and $B-V = -0.1$~mag;
Cernis 1993: SpT = B3III; Andersson et al.\ 2000: SpT = O9.5V--B0V;
the latter authors suggested that the IRAS ring, which is also visible
at radio wavelengths, is an H{\small{\sc{II}}} region associated with
the object). Our spectra, although noisy, show that the object is a
B-type rather than an F-type star. We therefore adopt the B3III
spectral type from Cernis (1993), who identified HD~278942 as a
`possible photometric B3III + F5I binary'. This choice implies a
radial velocity of $\sim$$31$~km~s$^{-1}$ (Table~\ref{tab:3}; despite
a repeatability error of $\sim$$16$~km~s$^{-1}$, the inferred radial
velocities of the two exposures are consistent at the level of
$\sim$$1$~km~s$^{-1}$). Using the SIMBAD/AGK3 B5 spectral type would
result in a radial velocity of $\sim$$38$~km~s$^{-1}$; Andersson's
classification would imply a radial velocity of
$\sim$70--90~km~s$^{-1}$.

\smallskip\noindent{\bf HD~281159:\/} We observed this star twice (in
subsequent exposures), obtaining $v_{\rm rad} = 9.1 \pm 1.8$ and
$7.8 \pm 2.1$~km~s$^{-1}$. Both SIMBAD and Duflot et al.\ (1995; the
WEB catalogue of radial velocities) list a literature radial velocity
of $+14 \pm 5$~km~s$^{-1}$, which seems in reasonable agreement at
first sight. There is, however, a long history behind these
values. B52, based on Moore's (1932) results, lists a radial velocity
of $+32$~km~s$^{-1}$ and sets a radial velocity variability
flag. Wilson \& Joy (1952) list `$+25$~km~s$^{-1}$ (variable radial
velocity)'. These authors based their mean value on 4 measurements:
$+42$, $+20$, $+80$, and $-43$~km~s$^{-1}$. The General Catalogue of
Radial Velocities (GCRV; 1953) lists $+25 \pm 5$~km~s$^{-1}$ and does
not mention velocity variability. Petrie \& Pearce (1961) list `$+7.0
\pm 4.7$~km~s$^{-1}$ (variable radial velocity)'. These authors based
their mean value on 6 measurements: $+34$, $-10$, $-3$, $-4$, $+19$,
and $+16$~km~s$^{-1}$. Evans (1967; `The revision of the GCRV') lists
$+14 \pm 5$~km~s$^{-1}$ based on 10 measurements. This mean value is
the number-of-measurements-weighted average of the mean results of
Wilson \& Joy (4 measurements) and Petrie \& Pearce (6
measurements). Evans did not copy any of the variability flags in the
literature, and his result ($+14 \pm 5$~km~s$^{-1}$) erroneously
suggests that the radial of HD~281159 is well-defined and
reliable. Evans's results were copied by both Duflot et al.\ (1995)
and SIMBAD. We thus conclude that the agreement between the results of
the latter two literature sources with our measurements is spurious
and we suspect that this object is an SB. As the radial velocity
determinations by Wilson \& Joy and Petrie \& Pearce have been
published without the associated epochs, we cannot analyse the data
for this star in the light of an SB model. Our mean radial velocity
(${\overline v}_{\rm rad} = 8.4 \pm 0.9$~km~s$^{-1}$) should be
treated with care.

\begin{figure}[t!] 
\centerline{\psfig{file=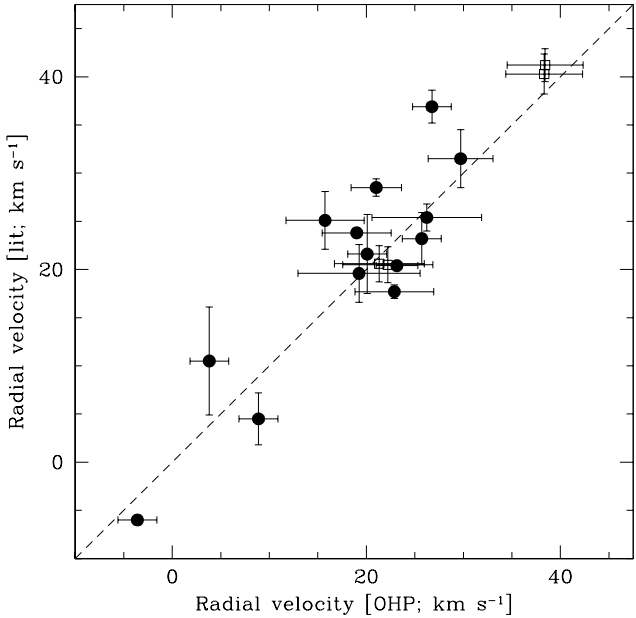,width=8.43truecm,silent=,clip=}} 
\caption[]{Comparison of OHP and literature radial velocities. The
filled circles denote 15 single stars (i.e., non-SBs). One of these,
HD~23268 with OHP and literature radial velocities of $4.1 \pm
0.6$ and $-20 \pm 8.5$~km~s$^{-1}$ (Duflot et al.\ 1995),
respectively, falls outside the plot. The four open squares refer to
the $2+2$ instantaneous measurements of the two single-lined SBs
HD~24190 ($v_{\rm rad} \sim 22$~km~s$^{-1}$) and 25799 ($v_{\rm rad}
\sim 38$~km~s$^{-1}$).}
\label{fig:6} 
\end{figure}

\subsection{External accuracy check} 
\label{subsec:external_accuracy} 
 
Our list of 29 stars with $v_{\rm rad}$ measurements contains the five
confirmed spectroscopic binaries discussed in Sect.~\ref{sec:sbs}, and
we have seen that our measured values are consistent with earlier
measurements. We were able to find reported radial velocity
measurements for 15 of the remaining 24 objects. The main sources, in
order of decreasing preference, are Zentelis (1983; Z83), Blaauw \&
van Albada (1963; BvA), Blaauw (1952; B52), and Grenier et al.\
(1999).

Figure~\ref{fig:6} shows our measured values ($v_{\rm OHP}$) for these
stars versus those in the literature ($v_{\rm lit}$). We have excluded
HD~23268. Its literature radial velocity from Duflot et al.\ (1995) of
$-20 \pm 8.5$~km~s$^{-1}$ is based on 3 measurements with unknown
source; it is consistent with our measurement of $4.1 \pm
0.6$~km~s$^{-1}$ at the $3$$\sigma$ level. There is reasonable
agreement between the literature values and our measurements. A small
offset from the diagonal can be expected, as we have set the zero of
our scale by the choice of standards, while in literature studies this
is done in various different ways (cf.\
Sect.~\ref{subsec:standard_stars}). We find that the weighted mean and
dispersion of $\Delta \equiv v_{\rm OHP} - v_{\rm lit}$ are $0.5$
and $3.3$~km~s$^{-1}$ (the straight mean and dispersion of $\Delta
\equiv (v_{\rm OHP} - v_{\rm lit}) / \sqrt{\sigma_{v, {\rm OHP}}^2 +
\sigma_{v, {\rm lit}}^2}$ are $0.1 \pm 0.9$, compared to the expected
value of $0 \pm 1$). These values are for the complete set of single
stars plus the $2+2$ SB1 exposures for HD~24190 and 25799
(Figure~\ref{fig:6}). We excluded HD~23268 and the three `outliers
above the diagonal' (HD~23060, 23478, and 24012) from this
sample. Removing the SB1 exposures from the sample results in 
$0.9$ and $3.5$~km~s$^{-1}$ (and $0.2 \pm 1.0$), while adding the
three outliers results in $-1.5$ and $5.4$~km~s$^{-1}$ (and
$-0.4 \pm 1.6$).

\begin{figure*}[t!] 
\centerline{\psfig{file=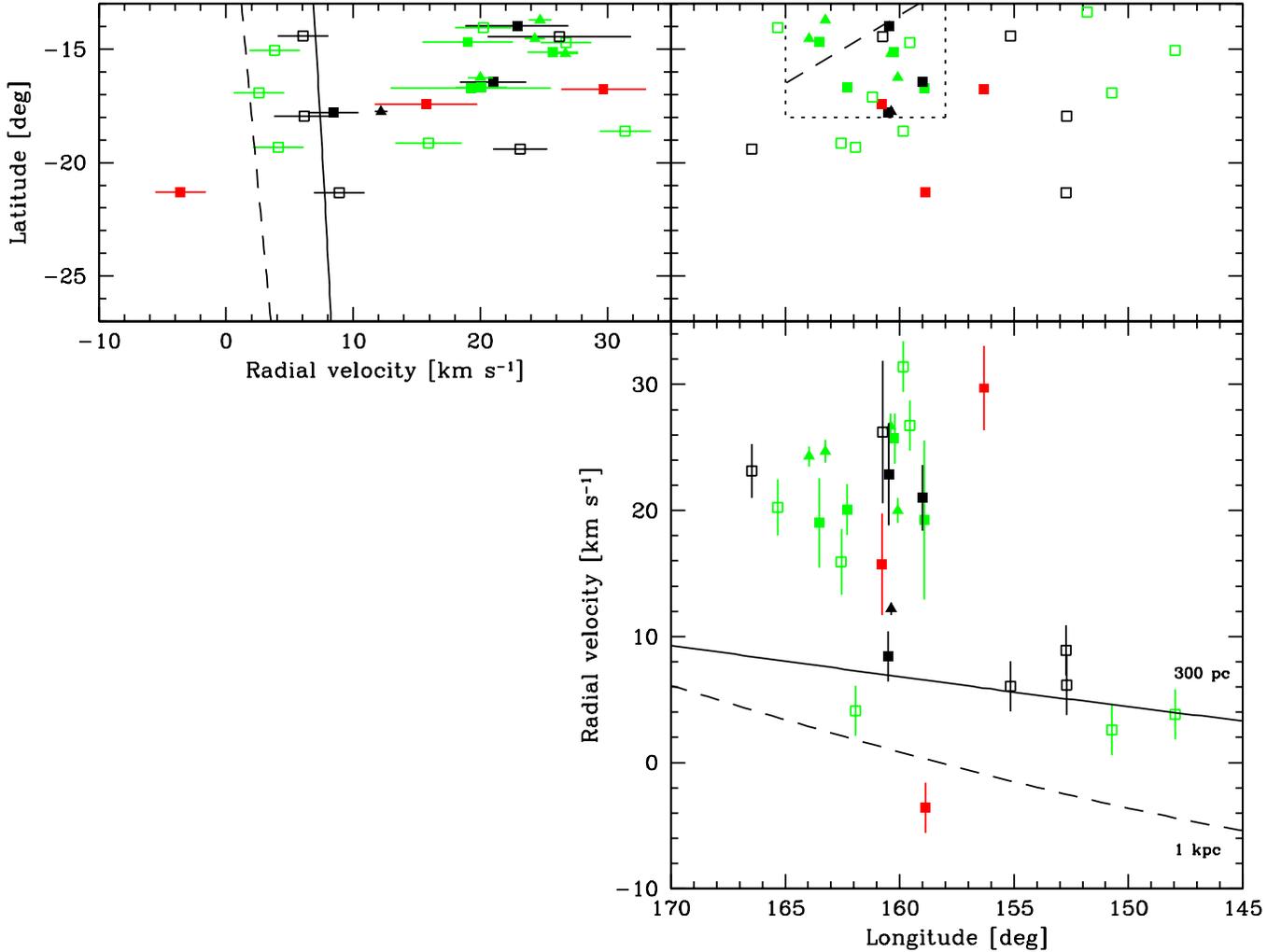,width=\textwidth,silent=,clip=}} 
\caption[]{The radial velocity measurements of our 29 targets as a
function of position on the sky. Squares denote our own radial
velocity measurements (24 stars), while triangles denote reported
systemic velocities for the five spectroscopic binaries
(Table~\ref{tab:6}). Open symbols refer to certain and possible
Hipparcos members (Z99), while filled symbols indicate 15 of the 17
classical Per~OB2 members from B52; significant overlap between these
groups exists (Sect.~\ref{subsec:B52}). The radial velocities of
HD~20987 and 23802 fall outside the radial-velocity panels (top left
and bottom right). The field of view (top right panel) is that used by
Z99 in their Hipparcos analysis of the Per~OB2 association. The dotted
box and dashed line on the sky denote Belikov et al.'s (2002b)
approximate extent of Per~OB2 and the separation between their alleged
subgroups a and b (see their figure~5). The lines in the
radial-velocity panels indicate the predicted radial velocities for
disk stars at 300~pc (full line) and 1~kpc (dashed line).}
\label{fig:summary} 
\end{figure*}

\section{Interpretation} 
\label{sec:interpretation} 
 
We now use our measurements, together with those for other certain and
proposed members of Per~OB2, to study the association. We first
discuss the distribution of radial velocities
(Sect.~\ref{subsec:dist}) and membership of individual stars
(Sect.~\ref{subsec:membership}), briefly address the internal
structure (Sect.~\ref{subsec:internal}), and then analyse the
colour-magnitude diagram (Sect.~\ref{subsec:cmd}).

\subsection{Distribution of radial velocities} 
\label{subsec:dist} 
 
We use our own measurements for 24 stars, and use the reported
systemic ($\gamma$) velocities for the five spectroscopic
binaries (Table~\ref{tab:6}) in our sample of candidate Per~OB2
members. The results are plotted in Figure~\ref{fig:summary}, which
shows the distribution of these objects on the sky, flanked by plots
of $v_{\rm rad}$ versus galactic longitude $\ell$ and latitude $b$. We
use different symbols for our new measurements (squares) and the
literature values (triangles).
 
The $v_{\rm rad}$ versus $\ell$ plot clearly shows a clump of stars
near $v_{\rm rad} \sim 23$~km~s$^{-1}$, containing many of the
classical Per~OB2 members (B52). The $v_{\rm rad}$ versus $b$ plot
shows a very similar separation. In addition to a few outliers, there
is a second clump of stars, covering a larger range in $\ell$, with a
small dispersion. The measured values coincide very nicely with the
expected $v_{\rm rad}$ for field stars in the direction of Per~OB2,
i.e., those obeying simple galactic rotation (in this direction,
however, most of the observed radial velocity is reflected Solar
motion). We conclude that these objects are unrelated field stars.
 
\begin{figure}[t!] 
\centerline{\psfig{file=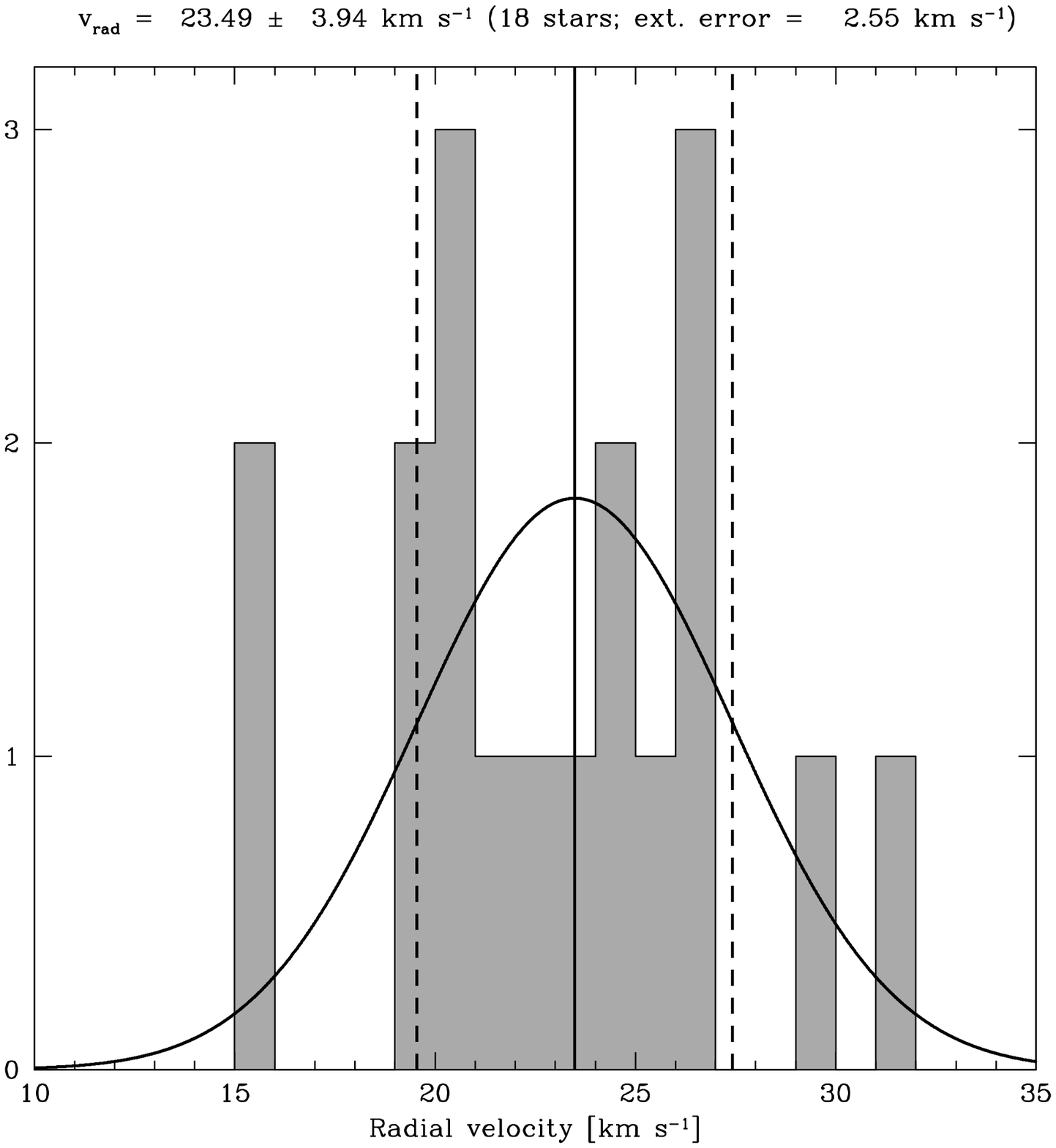,width=8.43truecm,silent=,clip=}} 
\caption[]{Radial velocity histogram of all 18 stars with OHP radial
velocities between 10 and 35~km~s$^{-1}$ (we excluded o~Per/HD~23180;
Sect.~\ref{subsec:membership}). We find a mean radial velocity of 
$23.5$~km~s$^{-1}$ (vertical line) and an associated dispersion of
 $3.9$~km~s$^{-1}$ (dashed vertical lines). The curve denotes a
reference Gaussian with a mean of $23.5$~km~s$^{-1}$ and 
$\sigma = 3.9$~km~s$^{-1}$. See Sect.~\ref{subsec:dist} for details.}
\label{fig:v_rad_histo} 
\end{figure}

The radial velocity separation between the association and the
Galactic disk allows us to determine the mean radial velocity of the
group. Our list of 29 stars contains 19 stars with radial velocities
between 10 and 35~km~s$^{-1}$. From this list, we reject the likely
non-member o~Per (HD~23180; see Sect.~\ref{subsec:membership}).
Figure~\ref{fig:v_rad_histo} shows the radial velocity histogram of
the 18 (candidate) members. From this histogram, we obtain a mean
radial velocity of $23.5$~km~s$^{-1}$ and an associated
dispersion $\sigma = 3.9$~km~s$^{-1}$. This mean velocity is
consistent with the value derived by Blaauw (1944), $19.4 \pm
1.7$~km~s$^{-1}$.

The dispersion of $3.9$~km~s$^{-1}$ amongst the measured radial
velocities in Per~OB2 provides an a posteriori external check of the
accuracy of our measurements, as it is an upper limit on this (the
measured dispersion arises from measurement errors, internal velocity
dispersion in the association, unrecognized duplicity, and/or the
presence of non-members in the sample). The internal dispersion is
probably only $1$--$3$~km~s$^{-1}$ (Z99), which suggests that the
external accuracy is $\sim$$3$~km~s$^{-1}$. This is in harmony with
the repeatability errors and the analysis of
Sect.~\ref{subsec:external_accuracy}.

\subsection{Membership} 
\label{subsec:membership} 

The clean separation of the association and the local disk stars in
radial velocity (Figure~\ref{fig:summary}) makes it possible to
improve some of the earlier membership assignments (notably B52 and
Z99). We discuss these in some detail here.

\subsubsection{Blaauw 1952 (B52)} 
\label{subsec:B52} 
 
B52 identified 17 stars as member of Per~OB2. Z99 confirmed eight of 
these: HIP~17313, 17735, 18081, 18111, 18246, 19039, 19178, and 19201 
(HD~22951, 23625, 24131, 24190, 24398, 25539, 25799, and 25833). Four 
of these eight are SBs (Table~\ref{tab:6}); their systemic radial 
velocities (from the literature) cluster around 
20--25~km~s$^{-1}$. The OHP radial velocities of the other four 
objects cluster in the same range. We thus conclude that all eight 
B52/Z99 Per~OB2 members are genuine members. 
 
Z99 identified four B52 members as possible astrometric member:
HIP~17387, 17448, 17465, and 18434 (HD~23060, 23180, 281159, and
24640). We observed all of them, and conclude that HD~23180 (o~Per), a
SB with a systemic velocity of $12.2$~km~s$^{-1}$
(Sect.~\ref{sec:sbs}), is most likely a non-member. According to their
spectral types and radial velocities, HD~23060 and 24640 are most
likely members. HD~281159 is possibly a SB and could be a member
(Sect.~\ref{sec:special}).
 
The remaining five B52 members are --- by definition --- Z99
non-members: HIP~16203, 16518, 17631, 18350, and 18614 (HD~21483,
21856, 23478, 24534, and 24912).

Both HD~21483 and 21856 were rejected by Z99. From the radial
velocities, we conclude that HD~21483 is clearly a non-member (this
object also has a deviating position on the sky). However, HD~21856
could be a member, certainly given its early B1V spectral type,
although its parallax is `small' ($1.99 \pm 0.82$~mas), its radial
velocity is `large' ($29.7 \pm 3.3$~km~s$^{-1}$), and its
(Hipparcos and Tycho-2) proper motion deviates from the mean of
Per~OB2.

HD~23478 (B3IV...) was not tested for membership in Z99 because the
Hipparcos astrometric data quality indicator was large (${\rm H}30 =
3.21$). Its OHP radial velocity ($15.8 \pm 4.0$~km~s$^{-1}$),
its parallax ($4.19 \pm 1.03$~mas), and its Hipparcos and Tycho-2
proper motions are consistent with membership.
 
HD~24912 is $\xi$~Per, a celebrated run-away star (Blaauw 1961;
Hoogerwerf, de Bruijne \& de Zeeuw 2001). It moves away from Per~OB2
with a relative radial velocity of $\sim$$40$~km~s$^{-1}$; we did not
observe this object.
 
The last of the remaining B52 members is X~Per (HD~24534). This O9.5pe
high-mass X-ray binary was identified as member by Blaauw (1944,
1952), although its absolute magnitude gave rise to doubts on this
classification. Z99 did not consider it because the Hipparcos proper
motion was of insufficient quality, and for this reason we did not
observe it. The reported systemic radial velocity is uncertain at the
level of $50$~km~s$^{-1}$ (B52; Wackerling 1972; Hutchings et al.\
1975; Duflot et al.\ 1995), so membership remains uncertain.

\subsubsection{De Zeeuw et al.\ 1999 (Z99)} 
\label{subsec:Z99} 

We observed 14 stars in addition to 15 of the 17 B52 members
(Sect.~\ref{subsec:B52}). All of these are certain or possible
astrometric Per OB2 members from Z99 (see Sect.~\ref{sec:obs}). Based
on their OHP radial velocities, we conclude that HD~18830, 19216,
19567, 20113, 22114, and 23268 are unrelated field stars in the disk
(`interlopers'). Based on their OHP radial velocities and the
Hipparcos data, we conclude that HD~24012, 24583, 24970, and 26499 are
members of Per~OB2. We briefly discuss the remaining four stars below.
 
\smallskip\noindent{\bf HD~20987:\/} The OHP radial velocity 
($-22.0$~km~s$^{-1}$) suggests that this Hipparcos acceleration
binary is a background field star. This suggestion is strengthened by
(i) the star's position on the sky $\sim$$10^{\circ\/}$ away from the
main body of the association; (ii) its relatively small parallax
($1.80 \pm 1.08$~mas); (iii) its early spectral type (B2V) combined
with its relatively faint magnitude ($V = 7.87$~mag).
 
\medskip\noindent{\bf HD~278942:\/} This faint object is a Hipparcos
component binary (Table~\ref{tab:1}). Cernis (1993) identified
HD~278942 as a `possible photometric B3III + F5I binary'. Its presence
in the Per~OB2 cloud and its parallax ($4.83 \pm 1.21$~mas) make it
clear that the object is roughly at the same distance as Per~OB2
($\sim$$318$~pc; Z99). We find a radial velocity of
$\sim$$38$~km~s$^{-1}$ for a B5 spectral type (SIMBAD/AGK3) and of
$\sim$$31$~km~s$^{-1}$ for a B3III spectral type (Cernis 1993; see
Sect.~\ref{sec:special} for details). The uncertain spectral type,
combined with duplicity, introduces a relatively large uncertainty in
the inferred radial velocity. We suspect that this object belongs to
Per~OB2.
 
\smallskip\noindent{\bf HD~23597:\/} Although the OHP radial velocity
is small ($\sim$$16$~km~s$^{-1}$), there is no compelling reason to
believe it is not a member of Per~OB2.

\smallskip\noindent{\bf HD~23802:\/} This B5Vn star is a Hipparcos
component binary with a parallax of $3.09 \pm 1.21$~mas. At first
sight, the OHP radial velocity ($\sim$$-50$~km~s$^{-1}$) seems
inconsistent with membership. The three OHP exposures, however, give a
relatively large spread in the inferred radial velocity
($\sim$$12$~km~s$^{-1}$), which, if significant, might be interpreted
as resulting from duplicity. This in turn could imply that the object
does belong to Per~OB2. Indirect evidence in favour of the above
arguments is provided by the highly negative value of the OHP radial
velocity: the correspondingly inferred distance of the object
($>$$4$~kpc) is inconsistent with its spectral type and magnitude.

\subsubsection{Literature radial velocities} 
\label{subsubsec:extra}

The analysis in Sections~\ref{subsec:B52}--\ref{subsec:Z99} leaves
three Z99 certain members which we did not observe, but for which
radial velocities are available from SIMBAD and BvA. The measurement
for HD~23244 (A0) is highly uncertain ($v_{\rm rad} = 15 \pm
12.4$~km~s$^{-1}$ from SIMBAD), but is formally consistent with
membership. HD~20825 (G5III) is a clear non-member at $6.2 \pm
1.2$~km~s$^{-1}$. HD~281157 (B3V; a Hipparcos component binary) has a
radial velocity of $20.9 \pm 2.1$~km~s$^{-1}$ from BvA, and hence is a
member.
 
\subsubsection{Conclusion on membership} 
 
It follows from Sections~\ref{subsec:B52}--\ref{subsec:Z99} that some
of the classical/possible members are in fact field stars. This shows
that the proper motion selection, while good, is not sufficient (this
statement is particularly true for Perseus~OB2 as the relative motion
of the group is mostly along the radial direction/line of
sight). Indeed, Z99 estimated the fraction of interlopers in their
list of Hipparcos members, and concluded that for Per~OB2 at 318~pc,
this is 70--100\% over the range B9, A and later. These numbers agree
with the Tycho-2 membership analysis of Belikov et al.\ (2002b) and
also with the findings of this investigation.

\renewcommand{\tabcolsep}{3.5pt} 
\begin{table}
\caption[]{Membership and photometry of Per~OB2. We first list the 29
objects observed by us (cf.\ Table~\ref{tab:1}). Below the horizontal
line, we added two classical members from B52 that were not observed
(Sect.~\ref{subsec:B52}) and three members from Z99 for which
literature radial velocities exist
(Sect.~\ref{subsubsec:extra}). Columns:
(1) HD number;
(2) spectral type;
(3) classical membership assignment (B52; C: certain member; P:
    possible member; N: non-member);
(4) Hipparcos membership (Z99);
(5) membership based on radial velocity (mostly this work);
(6) final membership, based on columns~4 and 5; in the case labeled
    with an asterisk, also photometric information was used;
(7) Johnson $B$ magnitude (from Tycho);
(8) Johnson $V$ magnitude (from Tycho); and
(9) visual extinction $A_V$ from Str\"omgren or Johnson photometry
    (Sect.~\ref{subsec:cmd}).
We do not list photometric data for eight clear non-members
(Sect.~\ref{subsec:cmd}). Membership for the run-away star $\xi$~Per
(HD~24912) comes down to semantics; we have indicated it by the label
R.}
\begin{center} 
\begin{tabular}{rlccccrrr}
\hline\hline
HD & SpT & \multicolumn{4}{c}{Membership} & $B$ & $V$ & $A_V$\\
(1) & (2) & (3) & (4) & (5) & (6) & (7) & (8) & (9)\\
\hline
 18830& A0       &  & C& N& \phantom{$^\ast$}N\phantom{$^\ast$}&      &       &       \\
 19216& B9V      &  & P& N& \phantom{$^\ast$}N\phantom{$^\ast$}&      &       &       \\
 19567& B9       &  & C& N& \phantom{$^\ast$}N\phantom{$^\ast$}&      &       &       \\
 20113& B8       &  & P& N& \phantom{$^\ast$}N\phantom{$^\ast$}&      &       &       \\
 20987& B2V      &  & C& N& \phantom{$^\ast$}N\phantom{$^\ast$}&      &       &       \\
 21483& B3III    & P& N& N& \phantom{$^\ast$}N\phantom{$^\ast$}&      &       &       \\
 21856& B1V      & C& N& C& \phantom{$^\ast$}C\phantom{$^\ast$}& 5.816&  5.899&  0.569\\
 22114& B8Vp     &  & P& N& \phantom{$^\ast$}N\phantom{$^\ast$}&      &       &       \\
 22951& B0.5V    & C& C& C& \phantom{$^\ast$}C\phantom{$^\ast$}& 4.927&  4.975&  0.737\\
 23060& B2Vp     & C& P& C& \phantom{$^\ast$}C\phantom{$^\ast$}& 7.585&  7.531&  1.072\\
 23180& B1III    & P& P& N& \phantom{$^\ast$}N\phantom{$^\ast$}& 3.871&  3.855&  0.886\\
 23268& A0       &  & C& N& \phantom{$^\ast$}N\phantom{$^\ast$}&      &       &       \\
 23478& B3IV...  & C&  & C& \phantom{$^\ast$}C\phantom{$^\ast$}& 6.717&  6.688&  0.783\\
 23597& B8       &  & C& C& \phantom{$^\ast$}C\phantom{$^\ast$}& 8.262&  8.225&  0.795\\
 23625& B2.5V    & C& C& C& \phantom{$^\ast$}C\phantom{$^\ast$}& 6.598&  6.564&  0.852\\
 23802& B5Vn     &  & C& P& \phantom{$^\ast$}P\phantom{$^\ast$}& 7.555&  7.386&  1.029\\
 24012& B5       &  & C& C& \phantom{$^\ast$}C\phantom{$^\ast$}& 7.835&  7.850&  0.632\\
 24131& B1V      & C& C& C& \phantom{$^\ast$}C\phantom{$^\ast$}& 5.747&  5.784&  0.737\\
 24190& B2V      & C& C& C& \phantom{$^\ast$}C\phantom{$^\ast$}& 7.458&  7.449&  0.936\\
 24398& B1Ib     & C& C& C& \phantom{$^\ast$}C\phantom{$^\ast$}& 2.966&  2.883&  1.158\\
 24583& B7V      &  & P& C& \phantom{$^\ast$}C\phantom{$^\ast$}& 9.047&  9.002&  0.626\\
 24640& B1.5V    & C& P& C& \phantom{$^\ast$}C\phantom{$^\ast$}& 5.431&  5.489&  0.563\\
 24970& A0       &  & P& C& \phantom{$^\ast$}N$^\ast$& 7.589&  7.466&  0.643\\
 25539& B3V      & C& C& C& \phantom{$^\ast$}C\phantom{$^\ast$}& 6.873&  6.874&  0.762\\
 25799& B3V...   & C& C& C& \phantom{$^\ast$}C\phantom{$^\ast$}& 7.066&  7.032&  0.802\\
 25833& B5V:p    & C& C& C& \phantom{$^\ast$}C\phantom{$^\ast$}& 6.710&  6.720&  0.571\\
 26499& B9       &  & C& C& \phantom{$^\ast$}C\phantom{$^\ast$}& 9.205&  9.057&  1\phantom{.000}\\
278942& B3III    &  & C& P& \phantom{$^\ast$}P\phantom{$^\ast$}&10.307&  9.175&  4.750\\
281159& B5V      & C& P& P& \phantom{$^\ast$}P\phantom{$^\ast$}& 9.278&  8.681&  2.720\\
\hline
 24534& O9.5pe   & C&  & P& \phantom{$^\ast$}P\phantom{$^\ast$}& 6.862&  6.793&  1.569\\
 24912& O7.5Iab: & P& N& N& \phantom{$^\ast$}R\phantom{$^\ast$}& 4.022&  4.042&  0.859\\[5pt]
 20825& G5III    &  & C& N& \phantom{$^\ast$}N\phantom{$^\ast$}&      &       &       \\
 23244& A0       &  & C& P& \phantom{$^\ast$}P\phantom{$^\ast$}&      &       &       \\
281157& B3V      &  & C& C& \phantom{$^\ast$}C\phantom{$^\ast$}& 9.811&  9.177&  2.930\\
\hline\hline
\end{tabular}
\end{center} 
\label{tab:7}
\end{table} 

Table~\ref{tab:7} summarizes our membership assignment for Per~OB2,
based on proper motions (B52, Z99), radial velocities (mostly this
work), and photometry (Sect.~\ref{subsec:cmd}). Membership
derived from radial velocity data (column~5) is based on a
$\pm$$2$$\sigma$ criterion. The remaining stars are, by definition,
classified as non-members, with the exception of the following five
`special cases', all of which are provisionally classified as possible
members: HD~23802 and 281159, two suspected (spectroscopic) binaries,
HD~278942, a faint target with an uncertain spectral type and
low-quality spectra, and HD~24534 and 23244, which have highly
uncertain literature radial velocities. Final membership (column~6) is
based on combining Hipparcos/astrometric (Z99) and radial velocity
membership (columns~4 and 5) following the logical scheme ${\rm
Col.~4} + {\rm Col.~5} \longrightarrow {\rm Col.~6}$, where $C+C
\longrightarrow C$, $P+P \longrightarrow P$, $N+N \longrightarrow N$,
$C+P \longrightarrow P$, $C+N \longrightarrow N$ (astrometric
interloper), $P+C \longrightarrow C$ (astrometric binary), $P+N
\longrightarrow N$, $N+C \longrightarrow C$ (astrometric binary), $N+P
\longrightarrow N$ (combination not present); C, N, and P stand for
certain, non-, and possible member. In practice, final membership in
our sample is effectively the same as radial velocity membership.

\subsection{Internal structure} 
\label{subsec:internal}

B52 did not distinguish subgroups for Per~OB2, although he did find
subgroups for several other OB associations (cf.\ Blaauw
1964). However, Blaauw speculated that subgroups would be created as
the Per~OB2 association would evolve. Mirzoyan et al.\ (1999) claimed
to have found substructure and expansion in Per~OB2 with the Hipparcos
analysis of 17 bright stars in the association. Mirzoyan, using
Hipparcos Input Catalogue (HIC) radial velocities, found two
`subgroups': one centered around $+17.4$~km~s$^{-1}$ and one around
$+26$~km~s$^{-1}$. Belikov et al.\ (2002b) also presented evidence for
two subgroups. We see no evidence for subdivision of Per~OB2 in
Figure~\ref{fig:v_rad_histo}, but our sample is small. For the
particular case of Per~OB2, Hipparcos astrometric radial velocities
offer no viable alternative to study internal structure issues: as the
relative motion of the association is mostly along the line of sight,
Hipparcos astrometric radial velocities are expected to be accurate to
$\sim$$2.5$~km~s$^{-1}$ (Dravins et al.\ 1999), similar to the
accuracy of the spectroscopic radial velocities presented here.

\begin{figure*}[t!] 
\centerline{\psfig{file=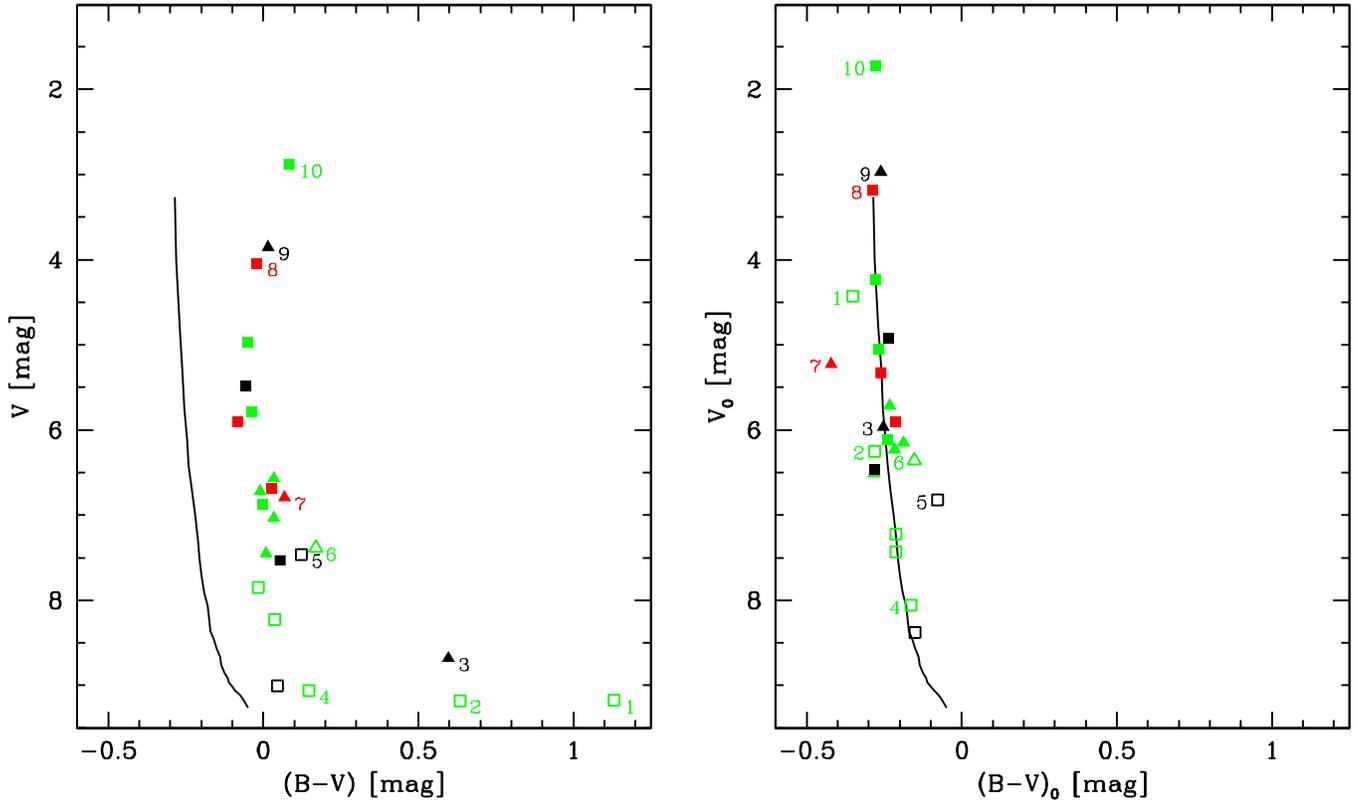,width=\textwidth,silent=,clip=}} 
\caption[]{Observed (left) and de-reddened (right) colour-magnitude
diagram of Per~OB2 ($R_V = 3.2$). The curves represent the zero-age
main sequence (Schmidt--Kaler 1982) at the mean distance of Per~OB2
($318$~pc; Z99). Symbols have the same meaning as in
Figure~\ref{fig:summary}, except that two new possible SBs have been
given triangular shapes: HD~23802 (labeled 6) and HD~281159 (labeled
3). Numerical labels (HD~numbers): 1 = 278942; 2 = 281157; 3 = 281159;
4 = 26499; 5 = 24970; 6 = 23802; 7 = 24534 (X~Per); 8 = 24912 (the
run-away $\xi$~Per); 9 = 23180 (o~Per); 10 = 24398 ($\zeta$~Per).  See
Sect.~\ref{subsec:cmd} for details.}
\label{fig:cmd} 
\end{figure*}
 
\subsection{Colour-magnitude diagram} 
\label{subsec:cmd} 
 
Figure~\ref{fig:cmd} shows the observed and de-reddened
colour-magnitude diagrams (CMDs). In order to construct these, we
removed the eight non-members identified above the horizontal line in
Table~\ref{tab:7} (denoted by `N' in column 6), but retained HD 24970
as this is a possible member based on the astrometry and radial
velocity measurement. We added the two classical members from B52 that
we did not observe ($\xi$~Per/HD~24912, which is a `member' in some
sense, and X~Per/HD~24534, which is a doubtful member;
Sect.~\ref{subsec:B52}) as well as HD~281157 which we classified as
member based on a literature radial velocity
(Sect.~\ref{subsubsec:extra}). We retrieved Johnson $B$ and $V$
magnitudes for the resulting 24 stars and derived individual
extinctions $A_V$ from both Str\"omgren and Johnson photometry using
$R_V = 3.2$. We adopted $A_V = 1$~mag, roughly the mean extinction for
Per~OB2 (e.g., Belikov et al.\ 2002b), for the star HD~26499 which
lacked appropriate photometry; this star is labeled `4' in
Figure~\ref{fig:cmd}.

The right panel of Figure~\ref{fig:cmd} displays a tight main
sequence. The stars HD~278942 and 281157 (labels `1' and `2',
respectively) both suffer from large extinctions ($A_V = 4.75$ and
$2.93$~mag, respectively). The accuracy of our simple extinction
correction is such that the de-reddened CMD locations are fully
consistent with membership. The locations of two B5V stars with
`peculiar'/deviating radial velocities (HD~281159 with $v_{\rm
rad} = 8.4$~km~s$^{-1}$ and HD~23802 with $v_{\rm rad} =
-52.3$~km~s$^{-1}$, labels `3' and `6', respectively) are also
consistent with membership, which suggests these stars could be
multiple. Our adoption of the average extinction to Per~OB2 for
HD~26499 (label `4', see above) puts it indeed on the dereddened main
sequence.

The CMD location of HD~24970 (label `5') raises doubt about its 
physical association with Per~OB2. Although the observed radial
velocity ($23.2 \pm 2.2$~km~s$^{-1}$) is consistent with
membership, we suspect that, given the relatively bright magnitude for
its spectral type (A0), its relatively `large' Hipparcos parallax
($4.95 \pm 0.99$~mas), and given the fact that this star is not a
certain but `only' a possible Z99 member, it is in fact a foreground
object (we found $A_V = 0.64$~mag, roughly the foreground extinction
to Per~OB2; Rydgren 1971). The peculiar location of X~Per can be
ascribed to the emission lines in its spectrum. The locations of the
runaway star $\xi$~Per (label `8') and the supergiant $\zeta$~Per
(label `10') are consistent with membership. Finally, o~Per (label
`9') seems to fit the zero-age main sequence, even though its radial
velocity indicates it is a non-member (the object is, nonetheless,
located in the Perseus molecular cloud).

While our radial velocity study has removed some interlopers from
earlier member lists of Per~OB2, including HD 21483 (B3III) and o~Per
(B1III), most of these have mid-B spectral types or later. As a
result, the location of the main sequence turn off is unchanged, and
so are the age determinations based on this.

\section{Concluding remarks} 
\label{sec:conc} 
 
We have presented radial velocity measurements derived from
high-resolution spectroscopy for 29 candidate B- and A-type members of
the association Per~OB2 obtained with AURELIE at Observatoire de
Haute--Provence. The radial velocities were found via cross
correlation with the spectra of observed standard stars with known
radial velocities. A careful observing strategy and data reduction,
rigorous removal of suspect exposures caused by an unstable detector,
and use of multiple standard stars to minimize template mismatch,
resulted in derived radial velocities accurate to
$\sim$$2$--$3$~km~s$^{-1}$. We have checked our error estimates by
comparison with literature measurements.
 
The new measurements, together with those for some additional
candidate members from earlier studies, show that Per~OB2 lies offset
by $\sim$$15$~km~s$^{-1}$ from the $v_{\rm rad}$ distribution for
field stars in this direction. The dispersion amongst the measurements
for the Per~OB2 members is only $3.9$~km~s$^{-1}$, which provides an
upper limit to our radial velocity accuracy. The internal
one-dimensional velocity dispersion of the association is smaller than
we can measure, but is likely to be $\sim$$1$--$3$~km~s$^{-1}$. 

The radial velocity measurements confirm many of the classical
astrometric members, but not all. Some earlier `secure' members turn
out to have discrepant $v_{\rm rad}$ values; this means they are
either previously unknown spectroscopic binaries or are genuine
non-members. The bulk of these discrepant stars have $v_{\rm rad}$ of
about $6$~km~s$^{-1}$, which puts them right at the velocity of the
disk (field) stars in this direction, suggesting they are
non-members. Removing these from the sample results in a very narrow
main sequence in the colour-magnitude diagram. The two outliers in
this diagram are the X-ray binary X Per, whose membership remains
uncertain, and the A0 star HD 24970. Even though its radial velocity
and proper motion are consistent with the space motion of Per OB2, we
conclude HD 24970 is a foreground object.

The classical study by BvA had already shown that a significant
fraction of the B-type members of Per~OB2 are spectroscopic binaries.
Our study has identified HD~23802, HD~281159, and possibly also the
peculiar object HD~278942, as additional candidate spectroscopic
binaries. It would be worthwhile to obtain multi-epoch spectroscopy
for all the B-type members of PerOB2, in order to derive accurate
orbits for the spectroscopic binaries, and possibly identify
additional ones. The effort required is significant, but would be
invaluable in determining the present day mass function of the
association, and the nature of the binary population.

The modest accuracy of the Hipparcos parallaxes and proper motions for
groups such as Per~OB2, combined with the relatively bright limiting
magnitude of the Hipparcos Catalogue, did not allow reliable
identification of members later than $\sim$B9--A0, even though these
are no doubt present (see, e.g., Preibisch et al.'s 2002 study of the
Sco--Cen OB association). A first step to probe membership of Per~OB2
for later spectral types was made by Belikov et al.\ (2002a, b), who
constructed a list of $\sim$$1000$ possible members of Per~OB2 based
on Tycho--2 proper motions and photometric information. Our results
demonstrate that a natural next step to distinguish members from
interlopers in this list would be to obtain spectra from which to
derive radial velocities (as well as spectral types and luminosity
classes).

ESA's future astrometry satellite GAIA (Perryman et al.\ 2001) will
extend proper motion studies to much fainter limiting magnitude than
Hipparcos, and will allow kinematic identification of low-mass
members, including all pre-main sequence objects, in Per~OB2 and other
nearby associations. GAIA will also allow study of more distant groups
in unprecedented detail. Besides astrometry, GAIA will also collect
multi-epoch spectroscopic measurements for all stars down to $V \sim
17$~mag. However, due to the specifics of the wavelength range used
($\sim$$250$~\AA\ around the Ca{\small{\sc II}}-triplet near 860~nm),
the resulting radial velocities for early-type stars will likely be of
modest precision. The currently planned ground-based all-sky
multi-object spectroscopic radial-velocity survey RAVE uses the same
spectral range, and may hence suffer from the same limitation. Our
study shows that dedicated (multi-epoch) spectroscopic studies of the
early-type stars in the nearby associations are feasible, and are
crucial in determining membership and the nature of the
population of high-mass stars in these young stellar groups.

\begin{acknowledgements} 
It is a pleasure to thank Adriaan Blaauw, Anthony Brown, Huib
Henrichs, Lex Kaper, and Werner Verschueren for many fruitful
discussions, and the referee, Jon Morse, for helpful comments.
The observations presented here were obtained at the Observatoire de
Haute Provence (CNRS) in France. This research has made use of the ADS
(NASA) and SIMBAD (CDS) services and the IRAF Data Reduction and
Analysis System. IRAF is distributed by the National Optical Astronomy
Observatories, which are operated by the Association of Universities
for Research in Astronomy, Inc., under cooperative agreement with the
National Science Foundation.
\end{acknowledgements}

\end{document}